# Unfounded Sets and Well-Founded Semantics
# of Answer Set Programs with Aggregates

**Mario Alviano**                                              ALVIANO@MAT.UNICAL.IT
**Francesco Calimeri**                                       CALIMERI@MAT.UNICAL.IT
**Wolfgang Faber**                                              FABER@MAT.UNICAL.IT
**Nicola Leone**                                               LEONE@MAT.UNICAL.IT
**Simona Perri**                                                PERRI@MAT.UNICAL.IT
*Department of Mathematics*
*University of Calabria*
*I-87030 Rende (CS), Italy*

## Abstract

Logic programs with aggregates ($LP^{\mathcal{A}}$) are one of the major linguistic extensions to Logic Programming (LP). In this work, we propose a generalization of the notions of unfounded set and well-founded semantics for programs with monotone and antimonotone aggregates ($LP^{\mathcal{A}}_{m,a}$ programs). In particular, we present a new notion of unfounded set for $LP^{\mathcal{A}}_{m,a}$ programs, which is a sound generalization of the original definition for standard (aggregate-free) LP. On this basis, we define a well-founded operator for $LP^{\mathcal{A}}_{m,a}$ programs, the fixpoint of which is called well-founded model (or well-founded set) for $LP^{\mathcal{A}}_{m,a}$ programs. The most important properties of unfounded sets and the well-founded semantics for standard LP are retained by this generalization, notably existence and uniqueness of the well-founded model, together with a strong relationship to the answer set semantics for $LP^{\mathcal{A}}_{m,a}$ programs. We show that one of the $\tilde{\mathcal{D}}$-well-founded semantics, defined by Pelov, Denecker, and Bruynooghe for a broader class of aggregates using approximating operators, coincides with the well-founded model as defined in this work on $LP^{\mathcal{A}}_{m,a}$ programs. We also discuss some complexity issues, most importantly we give a formal proof of tractable computation of the well-founded model for $LP^{\mathcal{A}}_{m,a}$ programs. Moreover, we prove that for general $LP^{\mathcal{A}}$ programs, which may contain aggregates that are neither monotone nor antimonotone, deciding satisfaction of aggregate expressions with respect to partial interpretations is coNP-complete. As a consequence, a well-founded semantics for general $LP^{\mathcal{A}}$ programs that allows for tractable computation is unlikely to exist, which justifies the restriction on $LP^{\mathcal{A}}_{m,a}$ programs. Finally, we present a prototype system extending DLV, which supports the well-founded semantics for $LP^{\mathcal{A}}_{m,a}$ programs, at the time of writing the only implemented system that does so. Experiments with this prototype show significant computational advantages of aggregate constructs over equivalent aggregate-free encodings.

## 1. Introduction

The use of *logical formulas* as a basis for a knowledge representation language was proposed about 50 years ago in some seminal works of McCarthy (1959), and McCarthy and Hayes (1969). However, it was soon realized that the monotonic nature of classical logic (the addition of new knowledge may only increase the set of consequences of a theory in classical logic) is not always suited to model commonsense reasoning, which sometimes is intrinsically nonmonotonic (Minsky, 1975). As an alternative, it was suggested to represent commonsense reasoning using logical languages with nonmonotonic consequence relations, which can better simulate some forms of human reasoning, allowing new knowledge to invalidate some of the previous conclusions. This observation opened a new and important research field, called *nonmonotonic reasoning*, and led to the definition and investigation of new logical formalisms, called *nonmonotonic logics*. The most popular nonmonotonic logics are circumscription (McCarthy, 1980, 1986), default logic (Reiter, 1980), and nonmonotonic modal logics (McDermott & Doyle, 1980; McDermott, 1982; Moore, 1985). Later on, from cross fertilizations between the field of nonmonotonic logics and that of logic programming, another nonmonotonic language, called Declarative Logic Programming (LP) has emerged, incorporating a nonmonotonic negation operator denoted by `not`. Declarative Logic Programming has gained popularity in the last years, and today it is a widely used formalism for knowledge representation and reasoning, with applications in various scientific disciplines and even in industry (Ricca, Alviano, Dimasi, Grasso, Ielpa, Iiritano, Manna, & Leone, 2010; Ricca, Grasso, Alviano, Manna, Lio, Iiritano, & Leone, 2011; Manna, Ricca, & Terracina, 2011; Manna, Ruffolo, Oro, Alviano, & Leone, 2011). In LP problems are solved by means of declarative specifications of requirements to be achieved. No ad-hoc algorithms are required.

Several semantics for LP have been proposed in the literature, which have to take care about the inherent non-monotonicity of the `not` operator in programs. The well-founded semantics (Van Gelder, Ross, & Schlipf, 1991) is one of the most prominent among them. It associates a three-valued model, the well-founded model, to every logic program. Originally, the well-founded semantics has been defined for normal logic programs, that is, standard logic programs with nonmonotonic negation. A distinguishing property of the well-founded semantics is that existence and uniqueness of the well-founded model is guaranteed for all logic programs. Moreover, the well-founded semantics is computable in polynomial time with respect to the input program in the propositional case.

Even if LP is a declarative programming language, standard LP does not allow for representing properties over sets of data in a natural way, a relevant aspect in many application domains. For addressing this insufficiency, several extensions of LP have been proposed, the most relevant of which is the introduction of aggregate functions (LP$^{\mathcal{A}}$; Kemp & Stuckey, 1991; Denecker, Pelov, & Bruynooghe, 2001; Dix & Osorio, 1997; Gelfond, 2002; Simons, Niemelä, & Soininen, 2002; Dell'Armi, Faber, Ielpa, Leone, & Pfeifer, 2003; Pelov & Truszczyński, 2004; Pelov, Denecker, & Bruynooghe, 2004). Among them, recursive definitions involving aggregate functions (i.e., aggregation in which aggregated data depend on the evaluation of the aggregate itself) are particularly interesting, as the definition of their semantics is not straightforward (Pelov, 2004; Faber, Leone, & Pfeifer, 2004; Son & Pontelli, 2007; Liu, Pontelli, Son, & Truszczynski, 2010). Note that a similar construct, referred to as *abstract constraint*, has been introduced in the literature (Marek & Truszczyński, 2004; Liu & Truszczyński, 2006; Son, Pontelli, & Tu, 2007; Truszczyński, 2010; Brewka, 1996). All of the results in this paper carry over also to LP with abstract constraints, for which well-founded semantics to our knowledge has not been defined so far.

In this paper we focus on the fragment of LP$^{\mathcal{A}}$ allowing for monotone and antimonotone aggregate expressions (LP$^{\mathcal{A}}_{m,a}$; Calimeri, Faber, Leone, & Perri, 2005). LP$^{\mathcal{A}}_{m,a}$ programs have many interesting properties. Among them, we highlight similarities between monotone aggregate expressions and positive standard literals, and between antimonotone aggregate expressions and negative standard literals. In particular, we take advantage of this aspect for defining unfounded sets and, based on this definition, a well-founded semantics for the $\mathrm{LP}_{m,a}^{\mathcal{A}}$ fragment. The well-founded semantics for $\mathrm{LP}_{m,a}^{\mathcal{A}}$ programs obtained in this way retains many desirable properties of the original well-founded semantics for LP, which it extends: For each $\mathrm{LP}_{m,a}^{\mathcal{A}}$ program a unique well-founded model exists, which is polynomial-time computable, approximates the programs answer sets, and coincides with the answer set on stratified $\mathrm{LP}_{m,a}^{\mathcal{A}}$ programs.

Actually it turns out that the well-founded semantics thus obtained coincides (on $\mathrm{LP}_{m,a}^{\mathcal{A}}$ programs) with a well-founded semantics proposed by Pelov, Denecker, and Bruynooghe (2007). Pelov et al. define several semantics of logic programs with aggregates using various approximating immediate consequence operators. The notion of logic program adopted by Pelov et al. is more general than the one considered in the present work, allowing for arbitrary first-order formulas in bodies, unrestricted aggregates, and non-Herbrand interpretations. Because of the equivalence of the two semantics, some properties proved by Pelov et al. carry over to this work as well. This applies to the results that the well-founded model is total on stratified programs (Theorem 9), that the well-founded model is contained in each answer set (Theorem 16), and that the well-founded model is computable in polynomial time (Theorem 21). However, the framework introduced in this article is considerably different from the one developed by Pelov et al., which allows for giving alternative proofs to these result. Vice versa, this article contains many new results, which carry over to the framework of Pelov et al. on $\mathrm{LP}_{m,a}^{\mathcal{A}}$ programs. In particular, it provides an alternative definition of the well-founded semantics, a characterization of answer sets by means of unfounded sets, and an implemented system computing the well-founded semantics, at the time of writing the only one of its kind.

We would like to point out that for most extensions of $\mathrm{LP}_{m,a}^{\mathcal{A}}$ programs that come to mind, the definition of unfounded sets would have to be considerably changed (see for instance the definition provided in Faber, 2005), and moreover the main desired properties of the well-founded semantics would no longer be guaranteed. For instance, the most obvious extension, including aggregate expressions that are neither monotone nor antimonotone would most likely not be computable in polynomial time: In fact, while the evaluation of aggregate expressions with respect to partial interpretations is tractable for monotone and antimonotone aggregates, the same task is coNP-complete for general aggregate expressions. Also, for instance allowing aggregates in rule heads would necessarily complicate the definition of unfounded sets, would not guarantee the existence of a well-founded model for every program, and would most likely not guarantee polynomial-time computability.

The concepts defined in this paper directly give rise to a computation method for the well-founded semantics on $\mathrm{LP}_{m,a}^{\mathcal{A}}$ programs. We have implemented this method, which is—to the best of our knowledge—the first of its kind. We have conducted experiments with this system on $\mathrm{LP}_{m,a}^{\mathcal{A}}$ encodings of a particular problem domain, and compared it with encodings not using aggregates. The latter encodings were tested with the system from which our prototype was derived and with XSB, a state-of-the-art system for computing the well-founded model. The experiments show a clear advantage of the $\mathrm{LP}_{m,a}^{\mathcal{A}}$ encodings run on our prototype system.

Summarizing, the main contributions of the paper are as follows.

- We define a new notion of unfounded set for logic programs with monotone and antimonotone aggregates ($LP_{m,a}^{\mathcal{A}}$ programs). This notion is a sound generalization of the concept of unfounded set previously given for standard logic programs. We show that our definition coincides with the original definition of unfounded sets (Van Gelder et al., 1991) on the class of normal (aggregate-free) programs, and that it shares its distinguishing properties (such as the existence of the greatest unfounded set).

- We define a well-founded operator $\mathcal{W}_{\mathcal{P}}$ for logic programs with aggregates, which extends the classical well-founded operator (Van Gelder et al., 1991). The total fixpoints of $\mathcal{W}_{\mathcal{P}}$ are exactly the answer sets of $\mathcal{P}$, and its least fixpoint $\mathcal{W}_{\mathcal{P}}^{\omega}(\emptyset)$ is contained in the intersection of all answer sets. We also show that the operator is equivalent to an operator defined by Pelov et al. (2007).

- We provide a declarative characterization of answer sets in terms of unfounded sets. In particular, we prove that the answer sets of an $LP_{m,a}^{\mathcal{A}}$ program are precisely the unfounded-free models.

- We show that reasoning with aggregates without restrictions may easily increase the complexity of the computation. In particular, we prove that deciding the truth or falsity of an aggregate expression with respect to a partial interpretation is a coNP-complete problem. However, while the problem is intractable in general, it is polynomial-time solvable for monotone and antimonotone aggregates.

- We analyze the complexity of the well-founded semantics, confirming and extending results in the work of Pelov et al. (2007). Importantly, it turns out that $\mathcal{W}_{\mathcal{P}}^{\omega}(\emptyset)$ is polynomial-time computable for propositional $LP_{m,a}^{\mathcal{A}}$ programs. For non-ground programs, the data-complexity remains polynomial, while the program complexity rises from P to EXPTIME, as for aggregate-free programs.

- We present a prototype system supporting the well-founded semantics defined in this article. The prototype, obtained by extending DLV, is the first system implementing a well-founded semantics for (unrestricted) $LP_{m,a}^{\mathcal{A}}$ programs.

- We report on experimental results on the implemented prototype. More specifically, we define the Attacks problem, a problem inspired by the classic Win-Lose problem often considered in the context of the well-founded semantics for standard logic programs. We compare the execution times of our prototype with an $LP_{m,a}^{\mathcal{A}}$ encoding and with equivalent LP encodings. In particular, one of the tested LP encodings is obtained by means of a compilation of aggregates into standard LP, which is also briefly presented in this paper. The obtained results evidence computational advantages for the problem encoding using aggregate expressions over those without them.

The presentation is organized as follows. In Section 2 we present the basics of the $LP^{\mathcal{A}}$ language and, in particular, we introduce the $LP_{m,a}^{\mathcal{A}}$ fragment. For this fragment, we define unfounded sets and well-founded semantics in Section 3. Relationships between well-founded semantics and answer set semantics are discussed in Section 4. A complexity analysis of the well-founded semantics for $LP_{m,a}^{\mathcal{A}}$ programs is reported in Section 5. In Section 6 we discuss the implemented prototype system and the experimentation. Finally, related work is discussed in Section 7, and in Section 8 we draw our conclusions.

## 2. The LP$^{\mathcal{A}}$ Language

Syntax, instantiation, interpretations and models of LP$^{\mathcal{A}}$ programs are introduced in this section. Moreover, we introduce the LP$^{\mathcal{A}}_{m,a}$ fragment of the language, for which we define a well-founded semantics in Section 3. For additional background on standard LP, we refer to the literature (Gelfond & Lifschitz, 1991; Baral, 2003).

### 2.1 Syntax

We assume sets of *variables*, *constants*, and *predicates* to be given. Similar to Prolog, we assume variables to be strings starting with uppercase letters and constants to be non-negative integers or strings starting with lowercase letters. Predicates are strings starting with lowercase letters. An *arity* (non-negative integer) is associated with each predicate. Moreover, the language allows for using built-in predicates (i.e., predicates with a fixed meaning) for the common arithmetic operations over positive integers (i.e., $=, \leq, \geq, +, \times$, etc.; written in infix notation), which are interpreted in the standard mathematical way.

#### 2.1.1 Standard Atom

A *term* is either a variable or a constant. A *standard atom* is an expression $p(t_1, \ldots, t_n)$, where $p$ is a *predicate* of arity $n$ and $t_1, \ldots, t_n$ are terms. An atom $p(t_1, \ldots, t_n)$ is ground if $t_1, \ldots, t_n$ are constants.

#### 2.1.2 Set Term

A *set term* is either a symbolic set or a ground set. A *symbolic set* is a pair $\{\mathit{Terms} : \mathit{Conj}\}$, where *Terms* is a list of terms (variables or constants) and *Conj* is a conjunction of standard atoms, that is, *Conj* is of the form $a_1, \ldots, a_k$ and each $a_i$ $(1 \leq i \leq k)$ is a standard atom. Intuitively, a set term $\{X : a(X, c), p(X)\}$ stands for the set of $X$-values making the conjunction $a(X, c), p(X)$ true, i.e., $\{X \mid a(X, c) \text{ and } p(X) \text{ are true}\}$. A *ground set* is a set of pairs of the form $\langle \mathit{consts} : \mathit{conj} \rangle$, where *consts* is a list of constants and *conj* is a conjunction of ground standard atoms.

#### 2.1.3 Aggregate Function

An *aggregate function* is of the form $f(S)$, where $S$ is a set term, and $f$ is an *aggregate function symbol*. Intuitively, an aggregate function can be thought of as a (possibly partial) function mapping multisets of constants to a constant. Throughout the remainder of the paper, we will adopt the notation of the DLV system (Leone, Pfeifer, Faber, Eiter, Gottlob, Perri, & Scarcello, 2006) for representing aggregates.

**Example 1** The most common aggregate functions are listed below:

- #min, minimal term, undefined for the empty set;
- #max, maximal term, undefined for the empty set;

- #count, number of terms;

- #sum, sum of integers;

- #times, product of integers;

- #avg, average of integers, undefined for the empty set.

### 2.1.4 Aggregate Atom

An *aggregate atom* is a structure of the form $f(S) \prec T$, where $f(S)$ is an aggregate function, $\prec \in \{<, \leq, >, \geq\}$ is a comparison operator, and $T$ is a term (variable or constant). An aggregate atom $f(S) \prec T$ is ground if $T$ is a constant and $S$ is a ground set.

**Example 2** The following are aggregate atoms in DLV notation:

$$\#\texttt{max}\{Z : r(Z), a(Z, V)\} > Y$$
$$\#\texttt{max}\{\langle 2 : r(2), a(2, m)\rangle, \langle 2 : r(2), a(2, n)\rangle\} > 1$$

### 2.1.5 Literal

A *literal* is either (i) a standard atom, or (ii) a standard atom preceded by the *negation as failure* symbol not, or (iii) an aggregate atom. Two standard literals are complementary if they are of the form $a$ and not $a$, for some standard atom $a$. For a standard literal $\ell$, we denote by $\neg.\ell$ the complement of $\ell$. Abusing of notation, if $L$ is a set of standard literals, then $\neg.L$ denotes the set $\{\neg.\ell \mid \ell \in L\}$.

### 2.1.6 Program

A *rule* $r$ is a construct of the form

$$a := \ell_1, \ldots, \ell_m.$$

where $a$ is a standard atom, $\ell_1, \ldots, \ell_m$ are literals, and $m \geq 0$. The atom $a$ is referred to as the *head* of $r$, and the conjunction $\ell_1, \ldots, \ell_m$ as the *body* of $r$. If the body is empty ($m = 0$), then the rule is called *fact*. We denote the head atom by $H(r) = a$, and the set of body literals by $B(r) = \{\ell_1, \ldots, \ell_m\}$. Moreover, the set of positive standard body literals is denoted by $B^+(r)$, the set of negative standard body literals by $B^-(r)$, and the set of aggregate body literals by $B^{\mathcal{A}}(r)$. A rule $r$ is ground if $H(r)$ and all the literals in $B(r)$ are ground. A *program* is a set of rules. A program is ground if all its rules are ground.

### 2.1.7 Safety

A *local* variable of a rule $r$ is a variable appearing solely in sets terms of $r$; a variable of $r$ which is not local is *global*. A rule $r$ is *safe* if both the following conditions hold: (i) for each global variable $X$ of $r$ there is a positive standard literal $\ell \in B^+(r)$ such that $X$ appears in $\ell$; (ii) each local variable of $r$ appearing in a symbolic set $\{Terms : Conj\}$ also appears in $Conj$. Note that condition (i) is the standard safety condition adopted in LP to guarantee that the variables are range restricted (Ullman, 1989), while condition (ii) is specific for aggregates. A program is safe if all its rules are safe.

**Example 3** Consider the following rules:

$$p(X) :\!- q(X, Y, V), \ \#\mathtt{max}\{Z : r(Z), \ a(Z, V)\} > Y.$$
$$p(X) :\!- q(X, Y, V), \ \#\mathtt{sum}\{Z : r(X), \ a(X, S)\} > Y.$$
$$p(X) :\!- q(X, Y, V), \ \#\mathtt{min}\{Z : r(Z), \ a(Z, V)\} > T.$$

The first rule is safe, while the second is not because the local variable $Z$ violates condition (ii). Also the third rule is not safe, since the global variable $T$ violates condition (i).

## 2.2 Program Instantiation, Interpretations and Models

In Section 3 we define a well-founded semantics for a relevant class of $\mathrm{LP}^{\mathcal{A}}$ programs. The well-founded semantics is defined for ground programs, while programs with variables are associated with equivalent ground programs. In this section we introduce preliminary notions such as program instantiation, interpretations and models.

### 2.2.1 Universe and Base

Given an $\mathrm{LP}^{\mathcal{A}}$ program $\mathcal{P}$, the *universe* of $\mathcal{P}$, denoted by $U_{\mathcal{P}}$, is the set of constants appearing in $\mathcal{P}$. The *base* of $\mathcal{P}$, denoted by $B_{\mathcal{P}}$, is the set of standard atoms constructible from predicates of $\mathcal{P}$ with constants in $U_{\mathcal{P}}$.

### 2.2.2 Instantiation

A *substitution* is a mapping from a set of variables to $U_{\mathcal{P}}$. Given a substitution $\sigma$ and an $\mathrm{LP}^{\mathcal{A}}$ object *obj* (rule, set, etc.), we denote by *obj* $\sigma$ the object obtained by replacing each variable $X$ in *obj* by $\sigma(X)$. A substitution from the set of global variables of a rule $r$ (to $U_{\mathcal{P}}$) is a *global substitution* for $r$; a substitution from the set of local variables of a set term $S$ (to $U_{\mathcal{P}}$) is a *local substitution* for $S$. Given a set term without global variables $S = \{\mathit{Terms} : \mathit{Conj}\}$, the *instantiation of $S$* is the following ground set:

$$inst(S) = \{\langle \mathit{Terms}\,\sigma : \mathit{Conj}\,\sigma\rangle \mid \sigma \text{ is a local substitution for } S\}.$$

A *ground instance* of a rule $r$ is obtained in two steps: First, a global substitution $\sigma$ for $r$ is applied, and then every set term $S$ in $r\sigma$ is replaced by its instantiation $inst(S)$. The instantiation $Ground(\mathcal{P})$ of a program $\mathcal{P}$ is the set of instances of all the rules in $\mathcal{P}$.

**Example 4** Consider the following program $\mathcal{P}_1$:

$$q(1) :\!- \mathtt{not}\ p(2,2). \qquad q(2) :\!- \mathtt{not}\ p(2,1). \qquad t(X) :\!- q(X), \ \#\mathtt{sum}\{Y : p(X, Y)\} > 1.$$
$$p(2,2) :\!- \mathtt{not}\ q(1). \qquad p(2,1) :\!- \mathtt{not}\ q(2).$$

The instantiation $Ground(\mathcal{P}_1)$ of $\mathcal{P}_1$ is the following program:

$$q(1) :\!- \mathtt{not}\ p(2,2). \qquad q(2) :\!- \mathtt{not}\ p(2,1). \qquad t(1) :\!- q(1), \ \#\mathtt{sum}\{\langle 1 : p(1,1)\rangle, \ \langle 2 : p(1,2)\rangle\} > 1.$$
$$p(2,2) :\!- \mathtt{not}\ q(1). \qquad p(2,1) :\!- \mathtt{not}\ q(2). \qquad t(2) :\!- q(2), \ \#\mathtt{sum}\{\langle 1 : p(2,1)\rangle, \ \langle 2 : p(2,2)\rangle\} > 1.$$

### 2.2.3 Aggregate Function Domain

Given a set $X$, let $\overline{2}^X$ denote the set of all multisets over elements from $X$. The domain of an aggregate function is the set of multisets on which the function is defined. Without loss of generality, we assume that aggregate functions map to $\mathbb{Z}$ (the set of integers).

**Example 5** Let us look at common domains for the aggregate functions of Example 1: #count is defined over $\overline{2}^{U_{\mathcal{P}}}$, #sum and #times over $\overline{2}^{\mathbb{Z}}$, #min, #max and #avg over $\overline{2}^{\mathbb{Z}} \setminus \{\emptyset\}$.

### 2.2.4 INTERPRETATION

An *interpretation* $I$ for an LP$^{\mathcal{A}}$ program $\mathcal{P}$ is a consistent set of standard ground literals, that is, $I \subseteq B_{\mathcal{P}} \cup \neg.B_{\mathcal{P}}$ and $I \cap \neg.I = \emptyset$. We denote by $I^+$ and $I^-$ the set of standard positive and negative literals occurring in $I$, respectively. An interpretation $I$ is *total* if $I^+ \cup \neg.I^- = B_{\mathcal{P}}$, otherwise $I$ is *partial*. The set of all the interpretations of $\mathcal{P}$ is denoted by $\mathcal{I}_{\mathcal{P}}$. Given an interpretation $I$ and a standard literal $\ell$, the evaluation of $\ell$ with respect to $I$ is defined as follows: (i) if $\ell \in I$, then $\ell$ is true with respect to $I$; (ii) if $\neg.\ell \in I$, then $\ell$ is false with respect to $I$; (iii) otherwise, if $\ell \notin I$ and $\neg.\ell \notin I$, then $\ell$ is undefined with respect to $I$. An interpretation also provides a meaning to set terms, aggregate functions and aggregate literals, namely a multiset, a value, and a truth value, respectively. We first consider a total interpretation $I$. The evaluation $I(S)$ of a set term $S$ with respect to $I$ is the multiset $I(S)$ defined as follows: Let $S^I = \{\langle t_1, ..., t_n \rangle \mid \langle t_1, ..., t_n : Conj \rangle \in S$ and all the atoms in *Conj* are true with respect to $I\}$; $I(S)$ is the multiset obtained as the projection of the tuples of $S_I$ on their first constant, that is, $I(S) = [t_1 \mid \langle t_1, ..., t_n \rangle \in S^I]$. The evaluation $I(f(S))$ of an aggregate function $f(S)$ with respect to $I$ is the result of the application of $f$ on $I(S)$.[1] If the multiset $I(S)$ is not in the domain of $f$, then $I(f(S)) = \bot$ (where $\bot$ is a fixed symbol not occurring in $\mathcal{P}$). A ground aggregate atom $\ell = f(S) \prec k$ is true with respect to $I$ if both $I(f(S)) \neq \bot$ and $I(f(S)) \prec k$ hold; otherwise, $\ell$ is false.

**Example 6** Let $I_1$ be a total interpretation having $I_1^+ = \{f(1), g(1, 2), g(1, 3), g(1, 4), g(2, 4), h(2), h(3), h(4)\}$. Assuming that all variables are local, we can check that:

- #count$\{X : g(X, Y)\} > 2$ is false; indeed, if $S_1$ is the corresponding ground set, then $S_1^{I_1} = \{\langle 1 \rangle, \langle 2 \rangle\}$, $I_1(S_1) = [1, 2]$ and #count$([1, 2]) = 2$.

- #count$\{X, Y : g(X, Y)\} > 2$ is true; indeed, if $S_2$ is the corresponding ground set, then $S_2^{I_1} = \{\langle 1, 2 \rangle, \langle 1, 3 \rangle, \langle 1, 4 \rangle, \langle 2, 4 \rangle\}$, $I_1(S_2) = [1, 1, 1, 2]$ and #count$([1, 1, 1, 2]) = 4$.

- #times$\{Y : f(X), g(X, Y)\} <= 24$ is true; indeed, if $S_3$ is the corresponding ground set, then $S_3^{I_1} = \{\langle 2 \rangle, \langle 3 \rangle, \langle 4 \rangle\}$, $I_1(S_3) = [2, 3, 4]$ and #times$([2, 3, 4]) = 24$.

- #sum$\{X : g(X, Y), h(Y)\} <= 3$ is true; indeed, if $S_4$ is the corresponding ground set, then $S_4^{I_1} = \{\langle 1 \rangle, \langle 2 \rangle\}$, $I_1(S_4) = [1, 2]$ and #sum$([1, 2]) = 3$.

- #sum$\{X, Y : g(X, Y), h(Y)\} <= 3$ is false; indeed, if $S_5$ is the corresponding ground set, then $S_5^{I_1} = \{\langle 1, 2 \rangle, \langle 1, 3 \rangle, \langle 1, 4 \rangle, \langle 2, 4 \rangle\}$, $I_1(S_5) = [1, 1, 1, 2]$ and #sum$([1, 1, 1, 2]) = 5$.;

- #min$\{X : f(X), h(X)\} >= 2$ is false; indeed, if $S_6$ is the corresponding ground set, then $S_6^{I_1} = \emptyset$, $I_1(S_6) = \emptyset$, and $I_1(\#$min$(\emptyset)) = \bot$ (we recall that $\emptyset$ is not in the domain of #min).

---

1. In this paper, we only consider aggregate functions value of which is polynomial-time computable with respect to the input multiset.

We now consider a partial interpretation $I$ and refer to an interpretation $J$ such that $I \subseteq J$ as an *extension* of $I$. If a ground aggregate atom $\ell$ is true (resp. false) with respect to *each* total interpretation $J$ extending $I$, then $\ell$ is true (resp. false) with respect to $I$; otherwise, $\ell$ is undefined.

**Example 7** Let $S_7$ be the ground set in the literal $\ell_1 = \#\mathtt{sum}\{\langle 1 : p(2,1)\rangle, \langle 2 : p(2,2)\rangle\} > 1$, and consider a partial interpretation $I_2 = \{p(2,2)\}$. Since each total interpretation extending $I_2$ contains either $p(2,1)$ or $\mathtt{not}\ p(2,1)$, we have either $I_2(S_7) = [2]$ or $I_2(S_7) = [1,2]$. Thus, the application of $\#\mathtt{sum}$ yields either $2 > 1$ or $3 > 1$, and thus $\ell_1$ is true with respect to $I_2$.

**Remark 1** Observe that our definitions of interpretation and truth values preserve "knowledge monotonicity": If an interpretation $J$ extends $I$ (i.e., $I \subseteq J$), each literal which is true with respect to $I$ is true with respect to $J$, and each literal which is false with respect to $I$ is false with respect to $J$ as well.

### 2.2.5 MODEL

Given an interpretation $I$, a rule $r$ is *satisfied* with respect to $I$ if at least one of the following conditions is satisfied: (i) $H(r)$ is true with respect to $I$; (ii) some literal in $B(r)$ is false with respect to $I$; (iii) $H(r)$ and some literal in $B(r)$ are undefined with respect to $I$. An interpretation $M$ is a *model* of an $\mathrm{LP}^{\mathcal{A}}$ program $\mathcal{P}$ if all the rules $r$ in $Ground(\mathcal{P})$ are satisfied with respect to $M$.

**Example 8** Consider again the program $\mathcal{P}_1$ of Example 4. Let $I_3$ be a total interpretation for $\mathcal{P}_1$ such that $I_3^+ = \{q(2), p(2,2), t(2)\}$. Then $I_3$ is a minimal model of $\mathcal{P}_1$.

## 2.3 The $\mathrm{LP}_{m,a}^{\mathcal{A}}$ Language

The definition of $\mathrm{LP}_{m,a}^{\mathcal{A}}$ programs, the fragment of $\mathrm{LP}^{\mathcal{A}}$ analyzed in this paper, is based on the following notion of monotonicity of literals.

### 2.3.1 MONOTONICITY

Given two interpretations $I$ and $J$, we say that $I \leq J$ if $I^+ \subseteq J^+$ and $I^- \supseteq J^-$. A ground literal $\ell$ is *monotone* if, for all interpretations $I, J$ such that $I \leq J$, we have that: (i) $\ell$ true with respect to $I$ implies $\ell$ true with respect to $J$, and (ii) $\ell$ false with respect to $J$ implies $\ell$ false with respect to $I$. A ground literal $\ell$ is *antimonotone* if the opposite happens, that is, for all interpretations $I, J$ such that $I \leq J$, we have that: (i) $\ell$ false with respect to $I$ implies $\ell$ false with respect to $J$, and (ii) $\ell$ true with respect to $J$ implies $\ell$ true with respect to $I$. A ground literal $\ell$ is *nonmonotone* if $\ell$ is neither monotone nor antimonotone. Note that positive standard literals are monotone, whereas negative standard literals are antimonotone. Aggregate literals, instead, may be monotone, antimonotone or nonmonotone. Some examples are shown below and the complete picture for the most common aggregate functions is summarized in Table 1.

**Example 9** Let us assume a universe in which all numerical constants are non-negative integers. All ground instances of the following aggregate literals are thus monotone:

Table 1: Character of the most common aggregate literals.

| Function | Domain | Operator | Character |
|---|---|---|---|
| #count | any | $>, \geq$ | monotone |
| | | $<, \leq$ | antimonotone |
| #sum | $\mathbb{N}$ | $>, \geq$ | monotone |
| | | $<, \leq$ | antimonotone |
| | $\mathbb{Z}$ | $<, \leq, >, \geq$ | *nonmonotone* |
| #times | $\mathbb{N}^+$ | $>, \geq$ | monotone |
| | | $<, \leq$ | antimonotone |
| | $\mathbb{N}, \mathbb{Z}$ | $<, \leq, >, \geq$ | *nonmonotone* |
| #min | any | $>, \geq$ | *nonmonotone**[*] |
| | | $<, \leq$ | monotone |
| #max | any | $>, \geq$ | monotone |
| | | $<, \leq$ | *nonmonotone**[*] |
| #avg | $\mathbb{N}, \mathbb{Z}$ | $<, \leq, >, \geq$ | *nonmonotone* |

[*] Antimonotone if the context guarantees that the set term of the aggregate never becomes empty.

$$\#\mathtt{count}\{Z : r(Z)\} > 1; \qquad \#\mathtt{sum}\{Z : r(Z)\} \geq 10.$$

Ground instances of the following literals are instead antimonotone:

$$\#\mathtt{count}\{Z : r(Z)\} < 1; \qquad \#\mathtt{sum}\{Z : r(Z)\} \leq 10.$$

### 2.3.2 $\mathrm{LP}_{m,a}^{\mathcal{A}}$ PROGRAMS

Let $\mathrm{LP}_{m,a}^{\mathcal{A}}$ denote the fragment of $\mathrm{LP}^{\mathcal{A}}$ allowing monotone and antimonotone literals. For an $\mathrm{LP}_{m,a}^{\mathcal{A}}$ rule $r$, the set of its monotone and antimonotone body literals are denoted by $B^m(r)$ and $B^a(r)$, respectively. An $\mathrm{LP}_{m,a}^{\mathcal{A}}$ program $\mathcal{P}$ is *stratified* if there exists a function $|| \cdot ||$, called *level mapping*, from the set of predicates of $\mathcal{P}$ to ordinals, such that for each pair $a$, $b$ of predicates, occurring in the head and body of a rule $r \in \mathcal{P}$, respectively: (i) if $b$ appears in an antimonotone literal, then $||b|| < ||a||$, (ii) otherwise $||b|| \leq ||a||$. Intuitively, stratification forbids recursion through antimonotone literals (for aggregate-free programs this definition coincides with the common notion of stratification with respect to negation).

**Example 10** Consider an $\mathrm{LP}_{m,a}^{\mathcal{A}}$ program consisting of the following rules:
$$q(X) :\!- p(X), \ \#\mathtt{count}\{Y : a(Y,X), b(X)\} \leq 2.$$
$$p(X) :\!- q(X), \ b(X).$$
and assume that the predicates $a$ and $b$ are defined by facts, which we do not include explicitly. The program is stratified, as the level mapping $||a|| = ||b|| = 1$, $||p|| = ||q|| = 2$ satisfies the required conditions. If we add the rule $b(X) :\!- p(X)$, then no such level-mapping exists, and the program becomes unstratified.

We would like to note that the definition of $\mathrm{LP}_{m,a}^{\mathcal{A}}$ could be enlarged, as in the form given above classifies literals independently of the context (that is, the program) in which they occur. Some aggregates that are nonmonotone by the definition given above, might not manifest their nonmonotone effects in a given context: If one limits the interpretations to be considered to those that do not violate the program in which the literal occurs, some interpretation pairs that violate monotonicity and antimonotonicity may no longer be present. In fact, one could refine the definition in this way (considering only pairs of non-violating interpretations of a given context program). The modified definition would enlarge the class of $\mathrm{LP}_{m,a}^{\mathcal{A}}$ programs, while retaining all of the results in this paper, but for simplicity of exposition we refrain from doing it formally. As an example, any aggregate atom involving #max with a < operator is formally not in $\mathrm{LP}_{m,a}^{\mathcal{A}}$, but when one considers occurrences in a program that has no non-violating interpretation $I$ such that $I(S) = \emptyset$ (where $S$ the set term of the aggregate), then the aggregate behaves in an antimonotone way in that particular program. We have noted these cases by a footnote in Table 1.

## 3. Unfounded Sets and Well-Founded Semantics

In this section we introduce a new notion of unfounded set for $\mathrm{LP}_{m,a}^{\mathcal{A}}$ programs, which extends the original definition for aggregate-free programs introduced by Van Gelder et al. (1991). Unfounded sets are then used for extending the well-founded semantics, originally defined for aggregate-free programs by Van Gelder et al., to $\mathrm{LP}_{m,a}^{\mathcal{A}}$ programs. We also highlight a number of desirable properties of this semantics. In the following we deal with ground programs, so we will usually denote by $\mathcal{P}$ a ground program. We will also use the notation $L \,\dot{\cup}\, \neg .L'$ for the set $(L \setminus L') \cup \neg .L'$, where $L$ and $L'$ are sets of standard ground literals.

**Definition 1 (Unfounded Set)** A set $\mathcal{X} \subseteq B_{\mathcal{P}}$ of ground atoms is an unfounded set for an $\mathrm{LP}_{m,a}^{\mathcal{A}}$ program $\mathcal{P}$ with respect to a (partial) interpretation $I$ if and only if, for each rule $r \in \mathcal{P}$ having $H(r) \in \mathcal{X}$, at least one of the following conditions holds:

(1) some (antimonotone) literal in $B^a(r)$ is false with respect to $I$, or

(2) some (monotone) literal in $B^m(r)$ is false with respect to $I \,\dot{\cup}\, \neg .\mathcal{X}$.

Intuitively, each rule with its head atom belonging to an unfounded set $\mathcal{X}$ is already satisfied with respect to $I$ (in case condition (1) holds), or it is satisfiable by taking as false all the atoms in the unfounded set (in case condition (2) holds). Note that, according to the definition above, the empty set is an unfounded set with respect to every program and interpretation.

**Example 11** Consider an interpretation $I_4 = \{a(1), a(2), a(3)\}$ for the following program $\mathcal{P}_2$:

$$
\begin{array}{ll}
r_1: & a(1) :\!\!- \; \#\mathtt{count}\{\langle 1\!:\!a(1)\rangle, \langle 2\!:\!a(2)\rangle, \langle 3\!:\!a(3)\rangle\} > 2. \\
r_2: & a(2). \\
r_3: & a(3) :\!\!- \; \#\mathtt{count}\{\langle 1\!:\!a(1)\rangle, \langle 2\!:\!a(2)\rangle, \langle 3\!:\!a(3)\rangle\} > 2.
\end{array}
$$

Then $\mathcal{X}_1 = \{a(1)\}$ is an unfounded set for $\mathcal{P}_2$ with respect to $I_4$, since condition (2) of Definition 1 holds for $r_1$ (the only rule with head $a(1)$). Indeed, the (monotone) literal appearing in $B^m(r_1)$ is false with respect to $I_4 \,\dot{\cup}\, \neg .\mathcal{X}_1 = \{\mathtt{not}\; a(1), a(2), a(3)\}$. Similarly, $\{a(3)\}$ and $\{a(1), a(3)\}$ are unfounded sets for $\mathcal{P}_2$ with respect to $I_4$. Clearly, also $\emptyset$ is an unfounded set. All other sets of atoms are not unfounded for $\mathcal{P}_2$ with respect to $I_4$.

As formalized below, Definition 1 generalizes the one given by Van Gelder et al. (1991) for aggregate-free programs: A set of standard atoms $\mathcal{X} \subseteq B_{\mathcal{P}}$ is an unfounded set for a program $\mathcal{P}$ with respect to an interpretation $I$ if and only if, for each rule $r \in \mathcal{P}$ such that $H(r) \in \mathcal{X}$, either (i) $B(r) \cap \neg.I \neq \emptyset$, or (ii) $B^+(r) \cap \mathcal{X} \neq \emptyset$.

**Theorem 1** For an aggregate-free program $\mathcal{P}$, Definition 1 is equivalent to the one introduced in the work of Van Gelder et al. (1991).

**Proof.** For an aggregate-free program $\mathcal{P}$, conditions (1) and (2) of Definition 1 are equivalent to (a) $B^-(r) \cap \neg.I \neq \emptyset$ and (b) $B^+(r) \cap \neg.(I \dot{\cup} \neg.\mathcal{X}) \neq \emptyset$, respectively. Condition (b) is equivalent to $B^+(r) \cap (\neg.(I \setminus \mathcal{X}) \cup \neg.\neg.\mathcal{X}) \neq \emptyset$, which holds if and only if either (b.1) $B^+(r) \cap \neg.(I \setminus \mathcal{X}) \neq \emptyset$, or (b.2) $B^+(r) \cap \mathcal{X} \neq \emptyset$. Condition (b.2) is exactly condition (ii) in the work of Van Gelder et al. Concerning condition (b.1), since $B^+(r)$ contains only positive literals, we can ignore the negative literals in $\neg.(I \setminus \mathcal{X})$, that is, the positive literals in $I \setminus \mathcal{X}$. By noting that the negative literals in $I \setminus \mathcal{X}$ are precisely the negative literals in $I$, we can then conclude that (b.1) is equivalent to $B^+(r) \cap \neg.I \neq \emptyset$. Finally, by combining the previous statement with condition (a) above, we obtain condition (i) in the work of Van Gelder et al. $\qquad \square$

Thus, Definition 1 is an alternative characterization of unfounded sets for aggregate-free programs. In fact, while condition (1) of Definition 1 does not exactly cover the first one in Van Gelder et al., condition (2) catches all cases of the second in the work of Van Gelder et al. and those "missed" by condition (1).

**Theorem 2** If $\mathcal{X}$ and $\mathcal{X}'$ are unfounded sets for an $\mathrm{LP}^{\mathcal{A}}_{m,a}$ program $\mathcal{P}$ with respect to an interpretation $I$, then $\mathcal{X} \cup \mathcal{X}'$ is an unfounded set for $\mathcal{P}$ with respect to $I$.

**Proof.** Let $r \in \mathcal{P}$ be such that $H(r) \in \mathcal{X} \cup \mathcal{X}'$. We want to show that either (1) some (antimonotone) literal in $B^a(r)$ is false with respect to $I$, or (2) some (monotone) literal in $B^m(r)$ is false with respect to $J = I \dot{\cup} \neg.(\mathcal{X} \cup \mathcal{X}')$. By symmetry, we can assume that $H(r)$ belongs to $\mathcal{X}$. Since $\mathcal{X}$ is an unfounded set with respect to $I$ by hypothesis, either (a) some (antimonotone) literal in $B^a(r)$ is false with respect to $I$, or (b) some (monotone) literal in $B^m(r)$ is false with respect to $K = I \dot{\cup} \neg.\mathcal{X}$. Case (a) is equals to (1). Thus, it remains to prove that case (b) implies (2). Indeed, we have that $J \leq K$ because $J^+ \subseteq K^+$ and $J^- \supseteq K^-$. Therefore, by definition of monotonicity, each monotone literal $\ell$ which is false with respect to $K$ is false with respect to $J$ as well, and so we are done. $\qquad \square$

As a corollary of Theorem 2, the union of all the unfounded sets is an unfounded set as well.

**Corollary 3** The union of all the unfounded sets for an $\mathrm{LP}^{\mathcal{A}}_{m,a}$ program $\mathcal{P}$ with respect to an interpretation $I$ is an unfounded set for $\mathcal{P}$ with respect to $I$ as well. We refer to this set as the *greatest unfounded set* of $\mathcal{P}$ with respect to $I$, denoted by $GUS_{\mathcal{P}}(I)$.

Below is an important monotonicity property of the greatest unfounded set.

**Proposition 4** Let $I$ and $J$ be interpretations for an $\mathrm{LP}^{\mathcal{A}}_{m,a}$ program $\mathcal{P}$. If $I \subseteq J$, then $GUS_{\mathcal{P}}(I) \subseteq GUS_{\mathcal{P}}(J)$.

**Proof.** Since $GUS_{\mathcal{P}}(J)$ is the union of all the unfounded sets for $\mathcal{P}$ with respect to $J$ by definition, it is enough to show that $\mathcal{X} = GUS_{\mathcal{P}}(I)$ is an unfounded set for $\mathcal{P}$ with respect to $J$. Thus, we want to show that, for each rule $r \in \mathcal{P}$ such that $H(r) \in \mathcal{X}$, either (1) some (antimonotone) literal in $B^a(r)$ is false with respect to $J$, or (2) some (monotone) literal in $B^m(r)$ is false with respect to $J \, \dot{\cup} \, \neg.\mathcal{X}$. We already know that $\mathcal{X}$ is an unfounded set for $\mathcal{P}$ with respect to $I$ by Corollary 3. Therefore, either (a) some (antimonotone) literal in $B^a(r)$ is false with respect to $I$, or (b) some (monotone) literal in $B^m(r)$ is false with respect to $I \, \dot{\cup} \, \neg.\mathcal{X}$. Since $I \subseteq J$, we have that $J$ and $J \, \dot{\cup} \, \neg.\mathcal{X}$ are extensions of the interpretations $I$ and $I \, \dot{\cup} \, \neg.\mathcal{X}$, respectively. Hence, by Remark 1, (a) implies (1) and (b) implies (2), and so we are done. □

We are now ready for extending the well-founded operator defined by Van Gelder et al. (1991) to the case of $\mathrm{LP}_{m,a}^{\mathcal{A}}$ programs.

**Definition 2** Let $\mathcal{P}$ be an $\mathrm{LP}_{m,a}^{\mathcal{A}}$ program. The *immediate logical consequence operator* $\mathcal{T}_{\mathcal{P}} : \mathcal{I}_{\mathcal{P}} \to 2^{B_{\mathcal{P}}}$ and the *well-founded operator* $\mathcal{W}_{\mathcal{P}} : \mathcal{I}_{\mathcal{P}} \to 2^{B_{\mathcal{P}} \cup \neg.B_{\mathcal{P}}}$ are defined as follows:

$$\mathcal{T}_{\mathcal{P}}(I) = \{\ell \in B_{\mathcal{P}} \mid \exists r \in \mathcal{P} \text{ such that } H(r) = \ell$$
$$\text{and all the literals in } B(r) \text{ are true with respect to } I\}$$
$$\mathcal{W}_{\mathcal{P}}(I) = \mathcal{T}_{\mathcal{P}}(I) \cup \; \neg.GUS_{\mathcal{P}}(I).$$

Intuitively, given an interpretation $I$ for a program $\mathcal{P}$, $\mathcal{W}_{\mathcal{P}}$ derives as true a set of atoms belonging to every model extending $I$ (by means of the $\mathcal{T}_{\mathcal{P}}$ operator). Moreover, $\mathcal{W}_{\mathcal{P}}$ derives as false all the atoms belonging to some unfounded set for $\mathcal{P}$ with respect to $I$ (by means of the $GUS_{\mathcal{P}}$ operator). Note that $\mathcal{T}_{\mathcal{P}}(I)$ and $GUS_{\mathcal{P}}(I)$ are set of atoms, so $\mathcal{W}_{\mathcal{P}}(I)^+ = \mathcal{T}_{\mathcal{P}}(I)$ and $\mathcal{W}_{\mathcal{P}}(I)^- = \neg.GUS_{\mathcal{P}}(I)$. The following proposition formalizes the intuition that Definition 2 extends the $\mathcal{W}_{\mathcal{P}}$ operator defined by Van Gelder et al. (1991) for standard programs to $\mathrm{LP}_{m,a}^{\mathcal{A}}$ programs.

**Proposition 5** Let $\mathcal{P}$ be an aggregate-free program. The $\mathcal{W}_{\mathcal{P}}$ operator of Definition 2 coincides with the $\mathcal{W}_{\mathcal{P}}$ operator defined by Van Gelder et al. (1991).

**Proof.** Since $\mathcal{W}_{\mathcal{P}}$ is equal to the union of $\mathcal{T}_{\mathcal{P}}$ and $\neg.GUS_{\mathcal{P}}$ in both cases, we have just to show that our definitions of $\mathcal{T}_{\mathcal{P}}$ and $GUS_{\mathcal{P}}$ coincide with those introduced by Van Gelder et al. (1991) for aggregate-free programs.

- The two immediate logical consequence operators ($\mathcal{T}_{\mathcal{P}}$) coincide for an aggregate-free program $\mathcal{P}$. Indeed, for each rule $r \in \mathcal{P}$, $B(r)$ has only standard literals.

- Our definition of $GUS_{\mathcal{P}}(I)$ coincides with the one of Van Gelder et al. (1991) for an aggregate-free program $\mathcal{P}$ and an interpretation $I$. Indeed, in both cases $GUS_{\mathcal{P}}(I)$ is defined as the union of all the unfounded sets for $\mathcal{P}$ with respect to $I$, and our notion of unfounded set coincides with the one in the work of Van Gelder et al. for standard programs by Theorem 1. □

We next show that a fixpoint of the well-founded operator $\mathcal{W}_{\mathcal{P}}$ is a (possibly partial) model.

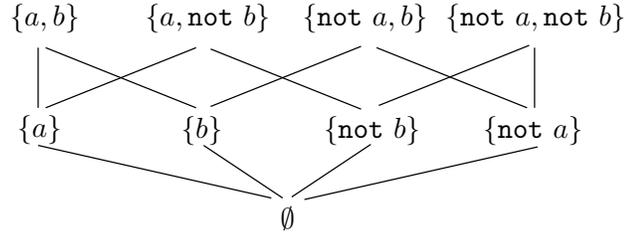

Figure 1: A meet semilattice

**Theorem 6** *Let $\mathcal{P}$ be an $\mathrm{LP}_{m,a}^{\mathcal{A}}$ program and $M$ a (partial) interpretation. If $M$ is a fixpoint of $\mathcal{W}_{\mathcal{P}}$, then $M$ is a (partial) model of $\mathcal{P}$.*

**Proof.** Let us assume that $\mathcal{W}_{\mathcal{P}}(M) = M$ holds. Thus, $\mathcal{T}_{\mathcal{P}}(M) \subseteq M$ and $\neg.GUS_{\mathcal{P}}(M) \subseteq M$ hold. Consider now a rule $r \in \mathcal{P}$. If all the literals in $B(r)$ are true with respect to $M$, then $H(r) \in \mathcal{T}_{\mathcal{P}}(M) \subseteq M$. If $H(r)$ is false with respect to $M$, then $H(r) \in GUS_{\mathcal{P}}(M)$. Since $GUS_{\mathcal{P}}(M)$ is an unfounded set for $\mathcal{P}$ with respect to $M$ by Corollary 3, either some literal in $B^{a}(r)$ is false with respect to $M$, or some literal in $B^{m}(r)$ is false with respect to $M \, \dot{\cup} \, \neg.GUS_{\mathcal{P}}(M) = M$. We can then conclude that $r$ is satisfied by $M$. □

The theorem below states that $\mathcal{W}_{\mathcal{P}}$ is a monotone operator in the meet semilattice induced on $\mathcal{I}_{\mathcal{P}}$ by the subset-containment relationship. We recall here that a meet semilattice is a partially ordered set which has a meet (or greatest lower bound) for any nonempty finite subset. An example of such a meet semilattice for a program with base $\{a, b\}$ is reported in Figure 1.

**Theorem 7** *Let $\mathcal{P}$ be an $\mathrm{LP}_{m,a}^{\mathcal{A}}$ program. The well-founded operator $\mathcal{W}_{\mathcal{P}}$ is a monotone operator in the meet semilattice $\langle \mathcal{I}_{\mathcal{P}}, \subseteq \rangle$.*

**Proof.** Since $\mathcal{W}_{\mathcal{P}}$ is equal to the union of $\mathcal{T}_{\mathcal{P}}$ and $\neg.GUS_{\mathcal{P}}$ by Definition 2, we have just to prove the monotonicity of the operators $\mathcal{T}_{\mathcal{P}}$ and $GUS_{\mathcal{P}}$.

- We first show that $\mathcal{T}_{\mathcal{P}}$ is a monotone operator, that is, for each pair of interpretations $I, J$ for $\mathcal{P}$ such that $I \subseteq J$, it holds that $\mathcal{T}_{\mathcal{P}}(I) \subseteq \mathcal{T}_{\mathcal{P}}(J)$. Consider an atom $\ell \in \mathcal{T}_{\mathcal{P}}(I)$. By Definition 2, there is a rule $r \in \mathcal{P}$ such that $H(r) = \ell$ and all the literals in $B(r)$ are true with respect to $I$. Since $I \subseteq J$, we can conclude that all the literals in $B(r)$ are true with respect to $J$ as well (see Remark 1), and so $H(r) = \ell$ belongs to $\mathcal{T}_{\mathcal{P}}(J)$ by Definition 2.

- We already know that $GUS_{\mathcal{P}}$ is a monotone operator from Proposition 4: For each pair of interpretations $I, J$ for $\mathcal{P}$ such that $I \subseteq J$, it holds that $GUS_{\mathcal{P}}(I) \subseteq GUS_{\mathcal{P}}(J)$.

□

We can now prove that the sequence $W_0 = \emptyset$, $W_{n+1} = \mathcal{W}_{\mathcal{P}}(W_n)$ is well-defined, that is, each element of the sequence is an interpretation.

**Theorem 8** *Let $\mathcal{P}$ be an $\mathrm{LP}_{m,a}^{\mathcal{A}}$ program. The sequence $W_0 = \emptyset$, $W_{n+1} = \mathcal{W}_{\mathcal{P}}(W_n)$ is well-defined.*

**Proof.** We use strong induction. The base case is trivial, since $W_0 = \emptyset$. In order to prove the consistency of $W_{n+1} = \mathcal{T}_{\mathcal{P}}(W_n) \cup \neg . GUS_{\mathcal{P}}(W_n)$, we assume the consistency of every $W_m$ such that $m \leq n$. Since $\mathcal{W}_{\mathcal{P}}$ is a monotone operator by Theorem 7, it is enough to show that $GUS_{\mathcal{P}}(W_n) \cap W_{n+1} = \emptyset$. To this end, we next show that any set $\mathcal{X}$ of atoms such that $\mathcal{X} \cap W_{n+1} \neq \emptyset$ is not an unfounded set for $\mathcal{P}$ with respect to $W_n$ (and so is not contained in $GUS_{\mathcal{P}}(W_n)$). Let $W_{m+1}$ be the first element of the sequence such that $\mathcal{X} \cap W_{m+1} \neq \emptyset$ (note that $m \leq n$). Consider any atom $\ell \in \mathcal{X} \cap W_{m+1}$. By definition of $\mathcal{T}_{\mathcal{P}}$, there is a rule $r \in \mathcal{P}$ having $H(r) = \ell$ and such that all the literals in $B(r)$ are true with respect to $W_m$. Note that no atom in $W_m$ can belong to $\mathcal{X}$ (for the way in which $W_{m+1}$ has been chosen). Thus, by Remark 1, all the literals in $B(r)$ are true with respect to both $W_n$ and $W_n \,\dot\cup\, \neg . \mathcal{X}$ (we recall that $W_n \supseteq W_m$ because $\mathcal{W}_{\mathcal{P}}$ is monotone). This ends the proof, as neither condition (1) nor (2) of Definition 1 hold for $\ell$. □

Theorem 8 and Theorem 7 imply that $\mathcal{W}_{\mathcal{P}}$ admits a least fixpoint (Tarski, 1955), which is referred to as the *well-founded model* of $\mathcal{P}$. The well-founded semantics of an $\mathrm{LP}_{m,a}^{\mathcal{A}}$ program $\mathcal{P}$ is given by this model. We can now state a first important property of the well-founded semantics of $\mathrm{LP}_{m,a}^{\mathcal{A}}$ programs.

**Property 1** For every $\mathrm{LP}_{m,a}^{\mathcal{A}}$ program, the well-founded model always exists and is unique.

Another important property of well-founded semantics easily follows from Proposition 5.

**Property 2** On aggregate-free programs, the well founded semantics as defined in this paper coincides with the classical well-founded semantics of Van Gelder et al. (1991).

Although the well-founded model, in general, might leave some atoms as undefined, there are cases where $\mathcal{W}_{\mathcal{P}}^{\omega}(\emptyset)$ is a total interpretation.

**Example 12** Consider the following program $\mathcal{P}_3$:
$$a(1) :- \#\texttt{sum}\{\langle 1 : a(1)\rangle, \langle 2 : a(2)\rangle\} > 2.$$
$$a(2) :- b.$$
$$b :- \texttt{not } c.$$

The iterated application of $\mathcal{W}_{\mathcal{P}}$ yields the following sets:

1. $\mathcal{W}_{\mathcal{P}}(\emptyset) = \{\texttt{not } a(1), \texttt{ not } c\};$
2. $\mathcal{W}_{\mathcal{P}}(\{\texttt{not } a(1), \texttt{ not } c\}) = \{\texttt{not } a(1), \texttt{ not } c, \ b\};$
3. $\mathcal{W}_{\mathcal{P}}(\{\texttt{not } a(1), \texttt{ not } c, \ b\}) = \{\texttt{not } a(1), \texttt{ not } c, \ b, \ a(2)\} \ = \mathcal{W}_{\mathcal{P}}^{\omega}(\emptyset).$

In this case, the well-founded model is total. Indeed, each atom in $B_{\mathcal{P}}$ is either true or false with respect to $\mathcal{W}_{\mathcal{P}}^{\omega}(\emptyset)$.

The totality of the well-founded model of the program above is due to its stratification, as formalized by the next theorem. Given Corollary 25, an equivalent result has been stated already by Pelov et al. (2007) as Theorem 7.2 and its Corollary 7.1. However, its proof is labelled as sketch by Pelov et al., which moreover relies on rather different formalisms than our proof.

**Theorem 9** On stratified $\text{LP}_{m,a}^{\mathcal{A}}$ programs, the well-founded model is total.

**Proof.** Let $\mathcal{P}$ be a stratified $\text{LP}_{m,a}^{\mathcal{A}}$ program. In order to prove that $\mathcal{W}_{\mathcal{P}}^{\omega}(\emptyset)$ is total, we show that each (standard) atom in $B_{\mathcal{P}} \setminus \mathcal{W}_{\mathcal{P}}^{\omega}(\emptyset)$ is false with respect to $\mathcal{W}_{\mathcal{P}}^{\omega}(\emptyset)$. By definition of stratification, there is a *level mapping* $|| \cdot ||$ of the (standard) predicates of $\mathcal{P}$ such that, for each pair $a$, $b$ of standard predicates occurring in the head and body of a rule $r \in \mathcal{P}$, respectively, the following conditions are satisfied: (i) if $b$ appears in an antimonotone literal, then $||b|| < ||a||$ holds; (ii) otherwise, if $b$ appears in a monotone literal, then $||b|| \leq ||a||$ holds. We are then in order to define a non-decreasing sequence of subsets of $B_{\mathcal{P}}$ as follows:

$$
\begin{aligned}
L_0 &= \emptyset \\
L_{i+1} &= L_i \cup \{\ell \in B_{\mathcal{P}} \mid \text{ the predicate of } \ell \text{ is } p \text{ and } ||p|| = i\}, \quad \forall i \in \mathbb{N}.
\end{aligned}
$$

Our aim is then to show that, for each $i \in \mathbb{N}$, the set $L_{i+1} \setminus \mathcal{W}_{\mathcal{P}}^{\omega}(\emptyset)$ is contained in $\neg.\mathcal{W}_{\mathcal{P}}^{\omega}(\emptyset)^-$. We use induction on $i$. The base case is trivial because $L_0 = \emptyset$ holds by definition. Now suppose that all the atoms in $L_i \setminus \mathcal{W}_{\mathcal{P}}^{\omega}(\emptyset)$ are false with respect to $\mathcal{W}_{\mathcal{P}}^{\omega}(\emptyset)$ in order to show that all the atoms in $L_{i+1} \setminus \mathcal{W}_{\mathcal{P}}^{\omega}(\emptyset)$ are false with respect to $\mathcal{W}_{\mathcal{P}}^{\omega}(\emptyset)$ as well. To this end, we prove that $L_{i+1} \setminus \mathcal{W}_{\mathcal{P}}^{\omega}(\emptyset)$ is an unfounded set for $\mathcal{P}$ with respect to $\mathcal{W}_{\mathcal{P}}^{\omega}(\emptyset)$. Consider a rule $r \in Ground(\mathcal{P})$ with $H(r) \in L_{i+1} \setminus \mathcal{W}_{\mathcal{P}}^{\omega}(\emptyset)$. We want to show that either (1) some (antimonotone) literal in $B^a(r)$ is false with respect to $\mathcal{W}_{\mathcal{P}}^{\omega}(\emptyset)$, or (2) some (monotone) literal in $B^m(r)$ is false with respect to $\mathcal{W}_{\mathcal{P}}^{\omega}(\emptyset) \dot{\cup} \neg.(L_{i+1} \setminus \mathcal{W}_{\mathcal{P}}^{\omega}(\emptyset))$. Since $H(r) \in L_{i+1}$, by definition of stratification the following propositions hold:

(a) each literal in $B^a(r)$ is either a negated standard atom belonging to $L_i$, or an aggregate literal depending only on atoms in $L_i$;

(b) each literal in $B^m(r)$ is either a standard atom belonging to $L_{i+1}$, or an aggregate literal depending only on atoms in $L_{i+1}$.

Since $H(r) \notin \mathcal{W}_{\mathcal{P}}^{\omega}(\emptyset)$ (that is, $H(r) \notin \mathcal{T}_{\mathcal{P}}(\mathcal{W}_{\mathcal{P}}^{\omega}(\emptyset))$), there is a literal $\ell$ in $B(r)$ such that $\ell$ is not true with respect to $\mathcal{W}_{\mathcal{P}}^{\omega}(\emptyset)$ (by definition of $\mathcal{T}_{\mathcal{P}}$). If $\ell$ is an antimonotone literal, we apply (a) and the induction hypothesis and conclude that (1) holds ($\ell$ cannot be undefined with respect to $\mathcal{W}_{\mathcal{P}}^{\omega}(\emptyset)$, so $\ell$ must be false). If $\ell$ is a monotone literal, we apply (b) and the induction hypothesis and conclude that (2) holds ($\ell$ cannot be undefined with respect to $\mathcal{W}_{\mathcal{P}}^{\omega}(\emptyset) \dot{\cup} \neg.(L_{i+1} \setminus \mathcal{W}_{\mathcal{P}}^{\omega}(\emptyset))$ and $\mathcal{W}_{\mathcal{P}}^{\omega}(\emptyset) \dot{\cup} \neg.(L_{i+1} \setminus \mathcal{W}_{\mathcal{P}}^{\omega}(\emptyset)) \leq \mathcal{W}_{\mathcal{P}}^{\omega}(\emptyset)$ holds, so $\ell$ must be false). $\qquad \square$

## 4. Answer Set Characterization via Unfounded Sets

The well-founded semantics is a three-valued semantics, that is, each program is associated with a model in which atoms are either true, false or undefined. Other semantics in the literature associate programs with two-valued models (i.e., models without undefined atoms). A commonly accepted two-value semantics in LP is the answer set semantics. In this section we present a number of results concerning unfounded sets and answer sets of $\text{LP}_{m,a}^{\mathcal{A}}$ programs. We first recall the definition of answer sets provided by Faber, Leone, and Pfeifer (2011).

**Definition 3 (Minimal Model)** A total model $M$ for $\mathcal{P}$ is (subset-)minimal if no total model $N$ for $\mathcal{P}$ exists such that $N^+ \subset M^+$. Note that, under these definitions, the words *interpretation* and *model* refer to possibly partial interpretations, while a *minimal model* is always a total interpretation.

We next provide the transformation by which the reduct of a ground program with respect to a total interpretation is formed. Note that this definition is a generalization (Faber et al., 2004) of the Gelfond-Lifschitz transformation (1991) for standard logic programs.

**Definition 4 (Program Reduct)** Given an LP$^{\mathcal{A}}$ program $\mathcal{P}$ and a total interpretation $I$, let $Ground(\mathcal{P})^I$ denote the transformed program obtained from $Ground(\mathcal{P})$ by deleting rules in which a body literal is false with respect to $I$, i.e.:

$$Ground(\mathcal{P})^I = \{r \in Ground(\mathcal{P}) \mid \text{ all the literals in } B(r) \text{ are true with respect to } I\}.$$

We are now ready for introducing the notion of answer set for LP$^{\mathcal{A}}$ programs.

**Definition 5 (Answer Set for LP$^{\mathcal{A}}$ Programs)** Given an LP$^{\mathcal{A}}$ program $\mathcal{P}$, a total interpretation $M$ of $\mathcal{P}$ is an answer set of $\mathcal{P}$ if and only if $M$ is a minimal model of $Ground(\mathcal{P})^M$.

**Example 13** Consider two total interpretations $I_5 = \{p(0)\}$ and $I_6 = \{\texttt{not } p(0)\}$ for the following two programs:

$$\mathcal{P}_4 = \{p(0) \coloneq \#\texttt{count}\{X : p(X)\} > 0.\}$$
$$\mathcal{P}_5 = \{p(0) \coloneq \#\texttt{count}\{X : p(X)\} \leq 0.\}$$

We then obtain the following transformed programs:

$Ground(\mathcal{P}_4)^{I_5} = Ground(\mathcal{P}_4) = \{p(0) \coloneq \#\texttt{count}\{\langle 0 : p(0)\rangle\} > 0.\}$
$Ground(\mathcal{P}_4)^{I_6} = \emptyset$
$Ground(\mathcal{P}_5)^{I_5} = \emptyset$
$Ground(\mathcal{P}_5)^{I_6} = Ground(\mathcal{P}_5) = \{p(0) \coloneq \#\texttt{count}\{\langle 0 : p(0)\rangle\} \leq 0.\}$

Hence, $I_6$ is the only answer set of $\mathcal{P}_4$. Indeed, $I_5$ is not a minimal model of $Ground(\mathcal{P}_4)^{I_5}$. Moreover, $\mathcal{P}_5$ has no answer sets. Indeed, $I_5$ is not a minimal model of $Ground(\mathcal{P}_5)^{I_5}$, and $I_6$ is not a model of $Ground(\mathcal{P}_5)^{I_6} = Ground(\mathcal{P}_5)$.

Note that any answer set $M$ of $\mathcal{P}$ is also a total model of $\mathcal{P}$ because $Ground(\mathcal{P})^M \subseteq Ground(\mathcal{P})$, and the rules in $Ground(\mathcal{P}) \setminus Ground(\mathcal{P})^M$ are satisfied with respect to $M$ (by Definition 4, each of these rules must have at least one body literal which is false with respect to $M$).

On the language LP$^{\mathcal{A}}_{m,a}$ considered in this work, answer sets as defined in Definition 5 coincide with stable models as defined by Pelov, Denecker, and Bruynooghe (2003) and hence also those defined by Pelov et al. (2007) and Son et al. (2007). This equivalence follows from Propositions 3.7 and 3.8 of Ferraris (2011), which respectively state that stable models of Pelov et al. (2003) on LP$^{\mathcal{A}}_{m,a}$ coincide with a semantics defined by Ferraris (2011), which in turn coincides with Definition 5 on a larger class of programs. This means that all our results involving answer sets also hold for these other semantics on LP$^{\mathcal{A}}_{m,a}$. On the other hand, this also implies that some of the results (for example Theorem 16) are consequences of results in the work of Pelov et al. (2007) by virtue of Theorem 24 in Section 7.

In the remainder of this section we highlight relevant relationships between answer sets and unfounded sets. Before introducing our results, let us provide an additional definition.

**Definition 6 (Unfounded-free Interpretation)** An interpretation $I$ for an $LP_{m,a}^{\mathcal{A}}$ program $\mathcal{P}$ is unfounded-free if and only if $I \cap \mathcal{X} = \emptyset$ holds for each unfounded set $\mathcal{X}$ for $\mathcal{P}$ with respect to $I$.

For total interpretations, an equivalent characterization of the unfounded-free property is given below.

**Lemma 10** A total interpretation $I$ for an $LP_{m,a}^{\mathcal{A}}$ program $\mathcal{P}$ is unfounded-free if and only if the empty set is the only subset of $I^+$ which is an unfounded set for $\mathcal{P}$ with respect to $I$.

**Proof.** ($\Rightarrow$) Straightforward: By Definition 6, $I$ is disjoint from all the unfounded set for $\mathcal{P}$ with respect to $I$.

($\Leftarrow$) We prove the contrapositive: If $I$ is not unfounded-free, then there exists a non-empty subset of $I^+$ which is an unfounded set for $\mathcal{P}$ with respect to $I$. From Definition 6, if $I$ is not unfounded-free, then there exists an unfounded set $\mathcal{X}$ for $\mathcal{P}$ with respect to $I$ such that $I \cap \mathcal{X} \neq \emptyset$. We next show that $I \cap \mathcal{X}$ is an unfounded set for $\mathcal{P}$ with respect to $I$, i.e., for each rule $r \in \mathcal{P}$ such that $H(r) \in I \cap \mathcal{X}$, either (1) some (antimonotone) literal in $B^a(r)$ is false with respect to $I$, or (2) some (monotone) literal in $B^m(r)$ is false with respect to $I \stackrel{.}{\cup} \neg.(I \cap \mathcal{X})$. Since $\mathcal{X}$ is an unfounded set, by Definition 1, either (a) some (antimonotone) literal in $B^a(r)$ is false with respect to $I$, or (b) some (monotone) literal in $B^m(r)$ is false with respect to $I \stackrel{.}{\cup} \neg.\mathcal{X}$. Thus, we can end the proof by showing that $I \stackrel{.}{\cup} \neg.\mathcal{X} = I \stackrel{.}{\cup} \neg.(I \cap \mathcal{X})$. To this end, observe that (i) $\neg.\mathcal{X} = \neg.(\mathcal{X} \setminus I) \cup \neg.(I \cap \mathcal{X})$. Moreover, since $I$ is total, $\neg.(B_{\mathcal{P}} \setminus I^+) = I^-$, and thus (ii) $\neg.(\mathcal{X} \setminus I) = \neg.(\mathcal{X} \setminus I^+) \subseteq I^- \subseteq I \setminus \mathcal{X}$. By using (i) in $I \stackrel{.}{\cup} \neg.\mathcal{X} = (I \setminus \mathcal{X}) \cup \neg.\mathcal{X}$ and simplifying with (ii) we obtain $I \stackrel{.}{\cup} \neg.\mathcal{X} = (I \setminus \mathcal{X}) \cup \neg.(I \cap \mathcal{X})$. We conclude by observing that $I \setminus \mathcal{X} = I \setminus (I \cap \mathcal{X})$, and thus $I \stackrel{.}{\cup} \neg.\mathcal{X} = I \stackrel{.}{\cup} \neg.(I \cap \mathcal{X})$ holds. □

Now we give another interesting characterization of total models for $LP_{m,a}^{\mathcal{A}}$ programs.

**Lemma 11** A total interpretation $M$ is a (total) model for an $LP_{m,a}^{\mathcal{A}}$ program $\mathcal{P}$ if and only if $\neg.M^-$ is an unfounded set for $\mathcal{P}$ with respect to $M$.

**Proof.** We start by observing that each rule $r \in \mathcal{P}$ such that $H(r) \in M^+$ is satisfied by $M$. Thus, we have to show that, for each rule $r \in \mathcal{P}$ with $H(r) \in \neg.M^-$, some literal in $B(r)$ is false with respect to $M$ if and only if either (1) some (antimonotone) literal in $B^a(r)$ is false with respect to $M$, or (2) some (monotone) literal in $B^m(r)$ is false with respect to $M \stackrel{.}{\cup} \neg.(\neg.M^-)$. To this end, it is enough to prove that $M \stackrel{.}{\cup} \neg.(\neg.M^-) = M$ holds. By definition, (∗) $M \stackrel{.}{\cup} \neg.(\neg.M^-) = (M \setminus \neg.M^-) \cup \neg.\neg.M^-$. From the consistency of $M$ we have that $M$ and $\neg.M^-$ are disjoint. Moreover, $\neg.\neg.M^- = M^-$ is a subset of $M$. By simplifying (∗) with the last two sentences, we obtain $M \stackrel{.}{\cup} \neg.(\neg.M^-) = M$. □

Now we give further characterizations of answer sets for $LP_{m,a}^{\mathcal{A}}$ programs.

**Theorem 12** A total model $M$ is an answer set of an $LP_{m,a}^{\mathcal{A}}$ program $\mathcal{P}$ if and only if $M$ is unfounded-free.

**Proof.** ($\Rightarrow$) We prove the contrapositive: If a total model $M$ of an $\text{LP}_{m,a}^{\mathcal{A}}$ program $\mathcal{P}$ is not unfounded-free, then $M$ is not an answer set of $\mathcal{P}$. By Lemma 10, since $M$ is a total interpretation and it is not unfounded-free, there exists an unfounded set $\mathcal{X}$ for $\mathcal{P}$ with respect to $M$ such that $\mathcal{X} \subseteq M^+$ and $\mathcal{X} \neq \emptyset$. Therefore, to prove that $M$ is not an answer set of $\mathcal{P}$, we next show that $M \;\dot{\cup}\; \neg.\mathcal{X}$ is a model of $\mathcal{P}^M$ such that $M \;\dot{\cup}\; \neg.\mathcal{X} \subset M$. To this end, consider a rule $r \in \mathcal{P}^M$. By Definition 4 of reduct, all the literals in $B(r)$ are true with respect to $M$, and so $H(r) \in M^+$ because $M$ is a model of $\mathcal{P}$ and $\mathcal{P}^M \subseteq \mathcal{P}$. We now have to consider two cases:

1. $H(r) \notin \mathcal{X}$. In this case, $H(r) \in M \;\dot{\cup}\; \neg.\mathcal{X}$ as well.

2. $H(r) \in \mathcal{X}$. In this case, since $\mathcal{X}$ is an unfounded set for $\mathcal{P}$ with respect to $M$, either (1) some literal in $B^a(r)$ is false with respect to $M$, or (2) some literal in $B^m(r)$ is false with respect to $M \;\dot{\cup}\; \neg.\mathcal{X}$. By previous considerations, since $r \in \mathcal{P}^M$, (1) cannot hold, and so we can conclude that some literal in $B(r)$ is false with respect to $M \;\dot{\cup}\; \neg.\mathcal{X}$.

Hence, we have that $r$ is satisfied by $M \;\dot{\cup}\; \neg.\mathcal{X}$ either by head (in case $H(r) \notin \mathcal{X}$), or by body (in case $H(r) \in \mathcal{X}$), and so we are done.

($\Leftarrow$) We prove the contrapositive: If a total model $M$ of an $\text{LP}_{m,a}^{\mathcal{A}}$ program $\mathcal{P}$ is not an answer set of $\mathcal{P}$, then $M$ is not unfounded-free. Since $M$ is a model of $\mathcal{P} \supseteq \mathcal{P}^M$ but not an answer set of $\mathcal{P}$, there exists a total model $N$ of $\mathcal{P}^M$ such that $N^+ \subset M^+$. We next show that $M^+ \setminus N^+$ is an unfounded set for $\mathcal{P}$ with respect to $M$, that is, for each rule $r \in \mathcal{P}$ such that $H(r) \in M^+ \setminus N^+$, either (1) some (antimonotone) literal in $B^a(r)$ is false with respect to $M$, or (2) some (monotone) literal in $B^m(r)$ is false with respect to $M \;\dot{\cup}\; \neg.(M^+ \setminus N^+)$.

We start by showing that $M \;\dot{\cup}\; \neg.(M^+ \setminus N^+) = N$. By definition, (a) $M \;\dot{\cup}\; \neg.(M^+ \setminus N^+) = (M \setminus (M^+ \setminus N^+)) \cup \neg.(M^+ \setminus N^+)$. From $N^+ \subset M^+$ we have (b) $M \setminus (M^+ \setminus N^+) = N^+ \cup M^-$. Moreover, since $N$ and $M$ are total interpretations and $N^+ \subset M^+$, we have (c) $N^- \supset M^-$ and (d) $\neg.(M^+ \setminus N^+) = N^- \setminus M^-$. Thus, by using (b) and (d) in (a) we obtain $M \;\dot{\cup}\; \neg.(M^+ \setminus N^+) = N^+ \cup M^- \cup (N^- \setminus M^-)$, and by observing that $M^- \cup (N^- \setminus M^-) = N^-$ holds because of (c), we conclude (e) $M \;\dot{\cup}\; \neg.(M^+ \setminus N^+) = N^+ \cup N^- = N$.

Consider now a rule $r \in \mathcal{P}$ such that $H(r) \in M^+ \setminus N^+$. We have to deal with two cases:

1. $r \in \mathcal{P} \setminus \mathcal{P}^M$. In this case, by Definition 4, there must be a literal $\ell \in B(r)$ such that $\ell$ is false with respect to $M$. If $\ell$ is an antimonotone literal, then (1) holds. Otherwise, $\ell$ is a monotone literal and so $\ell$ is false with respect to $N$ as well, since $N \leq M$; thus, (2) holds because of (e).

2. $r \in \mathcal{P}^M$. In this case, since $N$ is a model of $\mathcal{P}^M$ and $H(r)$ is false with respect to $N$ (because $H(r) \in M^+ \setminus N^+$ by assumption), there must be a literal $\ell \in B(r)$ such that $\ell$ is false with respect to $N$. If $\ell$ is an antimonotone literal, then $\ell$ is false with respect to $M$ as well, since $N \leq M$, and so (1) holds. Otherwise, $\ell$ is a monotone literal and (2) holds because of (e). $\qquad\square$

We are then ready to state an important connection between answer sets and unfounded sets.

**Theorem 13** A total interpretation $M$ for an $\text{LP}_{m,a}^{\mathcal{A}}$ program $\mathcal{P}$ is an answer set of $\mathcal{P}$ if and only if $GUS_{\mathcal{P}}(M) = \neg.M^-$.

**Proof.** ($\Rightarrow$) Let $M$ be an answer set of $\mathcal{P}$. By Lemma 11, $\neg.M^-$ is an unfounded set for $\mathcal{P}$ with respect to $M$, and hence $GUS_{\mathcal{P}}(M) \supseteq \neg.M^-$. By Theorem 12, $M$ is unfounded-free, and hence $GUS_{\mathcal{P}}(M) \subseteq \neg.M^-$ because $M$ is total. In sum, $GUS_{\mathcal{P}}(M) = \neg.M^-$.

($\Leftarrow$) Let $M$ be a total interpretation such that $GUS_{\mathcal{P}}(M) = \neg.M^-$. Then $M$ and $GUS_{\mathcal{P}}(M) = \neg.M^-$ are disjoint, and so $M$ is unfounded-free. Moreover, by Corollary 3, $GUS_{\mathcal{P}}(M) = M^-$ is an unfounded set for $\mathcal{P}$ with respect to $M$ and so, by applying Lemma 11, we conclude that $M$ is a model of $\mathcal{P}$. We are then in order to apply Theorem 12 ($M$ is an unfounded-free model of $\mathcal{P}$) and conclude that $M$ is an answer set of $\mathcal{P}$. $\square$

The following theorem shows that answer sets of $\mathrm{LP}^{\mathcal{A}}_{m,a}$ programs are exactly the total fixpoints of the well-founded operator defined in Section 3.

**Theorem 14** Let $M$ be a total interpretation for an $\mathrm{LP}^{\mathcal{A}}_{m,a}$ program $\mathcal{P}$. Then $M$ is an answer set for $\mathcal{P}$ if and only if $M$ is a fixpoint of the well-founded operator $\mathcal{W}_{\mathcal{P}}$.

**Proof.** ($\Rightarrow$) Let $M$ be an answer set of $\mathcal{P}$. We want to show that $M$ is a fixpoint of $\mathcal{W}_{\mathcal{P}}$, that is, $\mathcal{W}_{\mathcal{P}}(M) = M$. Our aim is then to show that $\mathcal{T}_{\mathcal{P}}(M) = M^+$ and $\neg.GUS_{\mathcal{P}}(M) = M^-$. Since $M$ is an answer set, by applying Theorem 13, we obtain $GUS_{\mathcal{P}}(M) = \neg.M^-$, which is equivalent to $\neg.GUS_{\mathcal{P}}(M) = M^-$. Therefore, it remains to prove that $\mathcal{T}_{\mathcal{P}}(M) = M^+$:

($\subseteq$) Consider an atom $\ell \in \mathcal{T}_{\mathcal{P}}(M)$. By Definition 2, there is a rule $r \in \mathcal{P}$ such that $H(r) = \ell$ and all the literals in $B(r)$ are true with respect to $M$. Thus, $\ell \in M^+$ holds because $M$ is a model of $\mathcal{P}$.

($\supseteq$) Consider an atom $\ell \in M^+$. Since $M$ is an answer set of $\mathcal{P}$, we can apply Theorem 12 and conclude that $M$ is unfounded-free. Hence, the (singleton) set $\{\ell\} \subseteq M^+$ is not an unfounded set for $\mathcal{P}$ with respect to $M$. Thus, by Definition 1, there is a rule $r \in \mathcal{P}$ such that $H(r) = \ell$ and neither (1) some (antimonotone) literal in $B^a(r)$ is false with respect to $M$, nor (2) some (monotone) literal in $B^m(r)$ is false with respect to $M \,\dot{\cup}\, \neg.\{\ell\}$. Since $M$ is a total interpretation, neither (1) nor (2) is equivalent to both (i) all the (antimonotone) literals in $B^a(r)$ are true with respect to $M$, and (ii) all the (monotone) literals in $B^m(r)$ are true with respect to $M \,\dot{\cup}\, \neg.\{\ell\}$. By observing that $M \,\dot{\cup}\, \neg.\{\ell\} \leq M$, we can state that (ii) implies that all the (monotone) literals in $B^m(r)$ are true with respect to $M$ as well. By combining the latter statement with (i) we obtain that all the literals in $B(r)$ are true with respect to $M$, and so $\ell \in \mathcal{T}_{\mathcal{P}}(M)$ by Definition 2.

($\Leftarrow$) Let $M$ be a total fixpoint of $\mathcal{W}_{\mathcal{P}}$, i.e., $\mathcal{W}_{\mathcal{P}}(M) = M$. Thus, $M^- = \neg.GUS_{\mathcal{P}}(M)$ by Definition 2, and so $M$ is an answer set for $\mathcal{P}$ because of Theorem 13. $\square$

Observe that Theorem 14 is a generalization of Theorem 5.4 of Van Gelder et al. (1991) to the class of $\mathrm{LP}^{\mathcal{A}}_{m,a}$ programs. It is also worth noting that $\mathcal{W}_{\mathcal{P}}(I)$ extends $I$ preserving its "correctness": If $I$ is contained in an answer set $M$, then $\mathcal{W}_{\mathcal{P}}(I)$ may add to $I$ some literals of $M$, but never introduces any literal which would be inconsistent with $M$.

**Proposition 15** Let $I$ be an interpretation for an $\mathrm{LP}^{\mathcal{A}}_{m,a}$ program $\mathcal{P}$, and let $M$ be an answer set for $\mathcal{P}$. If $I \subseteq M$, then $\mathcal{W}_{\mathcal{P}}(I) \subseteq M$.

**Proof.** This is a trivial consequence of the monotonicity of the operator $\mathcal{W}_\mathcal{P}$ (Theorems 7) and Theorem 14. Indeed, by Theorems 7, $\mathcal{W}_\mathcal{P}$ is $I \subseteq M$ implies $\mathcal{W}_\mathcal{P}(I) \subseteq \mathcal{W}_\mathcal{P}(M)$, and $\mathcal{W}_\mathcal{P}(M) = M$ by Theorem 14. □

We next show that the well-founded model of an $\mathrm{LP}_{m,a}^\mathcal{A}$ program is contained in all the answer sets (if any) of $\mathcal{P}$. We would like to point out that due to Theorem 24 in Section 7 (showing the equivalence of the well-founded operators defined in this work and the one defined in Pelov et al., 2007) and Propositions 3.77 and 3.8 of Ferraris (2011; showing the equivalence of answer sets in Faber et al., 2011 and stable models in Pelov et al., 2007), the following results also hold by virtue of the definitions of the well-founded and stable semantics in the work of Pelov et al., in particular due to Proposition 7.3 of that paper. We nevertheless also provide a proof using the concepts defined earlier.

**Theorem 16** Let $\mathcal{P}$ be an $\mathrm{LP}_{m,a}^\mathcal{A}$ program. For each answer set $M$ of $\mathcal{P}$, $\mathcal{W}_\mathcal{P}^\omega(\emptyset) \subseteq M$.

**Proof.** Let $M$ be an answer set of $\mathcal{P}$. Note that $\mathcal{W}_\mathcal{P}^\omega(\emptyset)$ is the limit of the sequence $W_0 = \emptyset$, $W_n = \mathcal{W}_\mathcal{P}(W_{n-1})$. We show that $W_n \subseteq M$ by induction on $n$. The base case is trivially true since $W_0 = \emptyset$ by definition. Now assume $W_n \subseteq M$ in order to show that $W_{n+1} \subseteq M$. Since $W_{n+1} = \mathcal{W}_\mathcal{P}(W_n)$ by definition and $W_n \subseteq M$ by induction hypothesis, we apply Proposition 15 and conclude that $W_{n+1} \subseteq M$. □

The theorem above suggests another property of well-founded semantics for $\mathrm{LP}_{m,a}^\mathcal{A}$ programs.

**Property 3** The well-founded semantics for $\mathrm{LP}_{m,a}^\mathcal{A}$ programs approximates the answer set semantics: The well-founded model is contained in the intersection of all answer sets (if any).

By combining Theorem 14 and Theorem 16, we obtain the following claim.

**Corollary 17** Let $\mathcal{P}$ be an $\mathrm{LP}_{m,a}^\mathcal{A}$ program. If $\mathcal{W}_\mathcal{P}^\omega(\emptyset)$ is a total interpretation, then it is the unique answer set of $\mathcal{P}$.

Therefore, by combining Theorem 9 and the corollary above, we obtain another property of well-founded semantics for $\mathrm{LP}_{m,a}^\mathcal{A}$ programs.

**Property 4** On stratified $\mathrm{LP}_{m,a}^\mathcal{A}$ programs, the well-founded model coincides with the unique answer set.

## 5. The Complexity of the Well-Founded Semantics

For the complexity analysis carried out in this section, we consider ground programs and polynomial-time computable aggregate functions (note that all example aggregate functions appearing in this paper fall into this class). However, we eventually provide a discussion on how results change when considering non-ground programs. We start with an important property of monotone and antimonotone aggregate literals.

**Lemma 18** Let $I$ be a partial interpretation for a ground $\mathrm{LP}_{m,a}^{\mathcal{A}}$ program $\mathcal{P}$. We define two total interpretations for $\mathcal{P}$ as follows: $I_{min} = I \cup \neg.(B_{\mathcal{P}} \setminus I)$ and $I_{max} = I \cup (B_{\mathcal{P}} \setminus \neg.I)$. For each (ground) aggregate literal $A$ occurring in $\mathcal{P}$, the following statements hold:

1. If $A$ is a monotone literal, then $A$ is true (resp. false) with respect to $I$ if and only if $A$ is true with respect to $I_{min}$ (resp. false with respect to $I_{max}$).

2. If $A$ is an antimonotone literal, then $A$ is true (resp. false) with respect to $I$ if and only if $A$ is true with respect to $I_{max}$ (resp. false with respect to $I_{min}$).

**Proof.** We start by noting that $I_{min}$ (resp. $I_{max}$) is a total interpretation extending $I$ and such that all the standard atoms which are undefined with respect to $I$ are false with respect to $I_{min}$ (resp. true with respect to $I_{max}$). Thus, we have $(*)$ $I_{min} \leq I \leq I_{max}$. If $A$ is monotone and true with respect to $I_{min}$ (resp. false with respect to $I_{max}$), then $A$ is true (resp. false) with respect to $I$ because of $(*)$. If $A$ is antimonotone and true with respect to $I_{max}$ (resp. false with respect to $I_{min}$), then $A$ is true (resp. false) with respect to $I$ because of $(*)$. We end the proof by observing that if $A$ is true (resp. false) with respect to $I$, then $A$ is true with respect to $I_{min}$ and $I_{max}$ by definition. $\square$

We are now ready to analyze the computational complexity of the well-founded semantics for $\mathrm{LP}_{m,a}^{\mathcal{A}}$ programs. Our analysis will lead to prove the following fundamental property.

**Property 5** The well-founded model for a ground $\mathrm{LP}_{m,a}^{\mathcal{A}}$ program is efficiently (polynomial-time) computable.

Given Corollary 25, this property also follows from Theorem 7.4 in the work of Pelov et al. (2007). In the following, we will provide an alternative proof based on the concepts defined earlier in this paper, which also leads to several interesting intermediate results.

Property 5 is not trivial because aggregates may easily increase the complexity of the evaluation. Indeed, even deciding the truth of an aggregate with respect to a partial interpretation is intractable in general; a similar observation has already been made by Pelov (2004). However, this task is polynomial-time computable for the aggregate literals occurring in $\mathrm{LP}_{m,a}^{\mathcal{A}}$ programs.

**Proposition 19** Deciding whether a ground aggregate literal $A$ is true (resp. false) with respect to a partial interpretation $I$ is:

(a) co-NP-complete in general;

(b) polynomial-time computable if $A$ is either a monotone or an antimonotone literal.

**Proof.** (a) As for the membership, we consider the complementary problem, that is, deciding whether a ground aggregate literal $A$ is not true (resp. not false) with respect to a partial interpretation $I$, and prove that it belongs to NP. In order to show that $A$ is not true (resp. not false) with respect to $I$ it is enough to find a total interpretation $J$ extending $I$ (that is, $J \supseteq I$) such that $A$ is false (resp. true) with respect to $J$. Thus, we can guess such a $J$ and check the falsity (resp. truth) of $A$ with respect to $J$ in polynomial time (if the aggregate function can be computed in polynomial time with respect to the size of the input multiset, as we are assuming).

As for the hardness, we first consider the problem of checking the truth of an aggregate and provide a polynomial-time reduction from TAUTOLOGY. The TAUTOLOGY problem is co-NP-complete and can be stated as follow: Given a proposition formula $\Phi$ on variables $X_1, \ldots, X_n$, does each truth assignment $v$ for the variables $X_1, \ldots, X_n$ satisfy the formula $\Phi$? Without loss of generality, we assume that $\Phi$ is a 3-DNF formula of the form

$$\Phi = D_1 \vee \cdots \vee D_m,$$

where each disjunct $D_i$ is a conjunction $\ell_i^1 \wedge \ell_i^2 \wedge \ell_i^3$, and each $\ell_i^j$ is a positive or negative literal (note that, in the context of TAUTOLOGY, the term "literal" denotes a variable $X_k$ or a variable preceded by the negation symbol $\neg$). For a given $\Phi$, we then consider a partial interpretation $I = \{\top\}$ and construct an aggregate literal $A = \#\mathtt{sum}\{S\} \geq 1$, where $S$ contains two groups of elements. The elements in the first group represent disjuncts of $\Phi$ and are

$$\langle 1 : \gamma(\ell_i^1), \gamma(\ell_i^2), \gamma(\ell_i^3) \rangle, \qquad i = 1, \ldots, m \,,$$

where, for each $i = 1, \ldots, m$ and $j = 1, \ldots, 3$, the propositional atom $\gamma(\ell_i^j)$ is defined as follows:

$$\gamma(\ell_i^j) = \begin{cases} x_k^t & \text{if } \ell_i^j \text{ is a positive literal } X_k, \text{ for some } k \in \{1, \ldots, n\}. \\ x_k^f & \text{if } \ell_i^j \text{ is a negative literal } \neg X_k, \text{ for some } k \in \{1, \ldots, n\}. \end{cases}$$

The elements in the second group represent variables of $\Phi$ and are as follows:

$$\begin{cases} \langle \ \ 1, \ x_k \colon \top \rangle \\ \langle -1, \ x_k \colon x_k^t \rangle \\ \langle -1, \ x_k \colon x_k^f \rangle \\ \langle \ \ 1, \ \hat{x_k} \colon x_k^t, \ x_k^f \rangle \end{cases} \,, \qquad k = 1, \ldots, n \,,$$

where $x_k$ and $\hat{x_k}$ are constants associated with the variable $X_k$. Note that, for each variable $X_k$ of $\Phi$, there are two atoms in $A$, $x_k^t$ and $x_k^f$. Thus, for each interpretation $J$, four cases are possible:

(1) $\{\mathtt{not}\ x_k^t, \ \mathtt{not}\ x_k^f\} \subseteq J$: In this case, only $\langle 1, \ x_k \colon \top \rangle$ contribute to the evaluation of $A$, and its contribution is 1;

(2) $\{x_k^t, \ x_k^f\} \subseteq J$: In this case, all the four elements contribute to the evaluation of $A$, and thus their contribution is $1 - 1 + 1 = 1$ (note that $\langle -1, \ x_k \colon x_k^t \rangle$ and $\langle -1, \ x_k \colon x_k^f \rangle$ give a total contribution of $-1$ because of our pure set approach);

(3) $\{x_k^t, \ \mathtt{not}\ x_k^f\} \subseteq J$: In this case, only $\langle 1, \ x_k \colon \top \rangle$ and $\langle -1, \ x_k \colon x_k^t \rangle$ contribute, giving $1 - 1 = 0$;

(4) $\{\mathtt{not}\ x_k^t, \ x_k^f\} \subseteq J$: In this case, only $\langle 1, \ x_k \colon \top \rangle$ and $\langle -1, \ x_k \colon x_k^f \rangle$ contribute, giving $1 - 1 = 0$.

Thus, for each $k \in \{1, \ldots, n\}$, the total contribution of the four elements of $S$ associated with the variable $X_k$ is either 0 or 1. Note that also the total contribution of the other elements of $S$ (i.e., those in the first group) is either 0 or 1. Therefore, if there is $k \in \{1, \ldots, n\}$ such that either case (1) or (2) occurs, the interpretation $J$ trivially satisfies $A$. Otherwise, $J$ is such that, for each variable $k \in \{1, \ldots, n\}$, either (3) or (4) occurs. In this case, we say that $J$ is a *good interpretation*.

We next define a one-to-one mapping between the set of assignments for $\Phi$ and the set of good interpretations. Let $v$ be an assignment for $\Phi$. The good interpretation $I_v$ associated with $v$ is such that $\top \in I_v$ and

$$\begin{cases} \{x_k^t, \; \texttt{not } x_k^f\} \subseteq I_v & \text{if } v(X_k) = 1 \\ \{\texttt{not } x_k^t, \; x_k^f\} \subseteq I_v & \text{if } v(X_k) = 0 \end{cases}, \qquad k = 1, \ldots, n \, .$$

We want to show that $v$ satisfies $\Phi$ if and only if $A$ is true with respect to $I_v$. Since $I_v$ is a good interpretation, the elements of $S$ in the second group give a total contribution of 0, and so we have just to consider the elements of $S$ in the first group. These elements give a contribution of 1 if and only if $\{\gamma(\ell_i^1), \gamma(\ell_i^2), \gamma(\ell_i^3)\} \subseteq I$ holds for at least one $i \in \{1, \ldots, n\}$, and this holds if and only $v(D_i) = 1$ holds for the disjunct $D_i$. We can then conclude that $A$ is true with respect to $I_v$ if and only $v(\Phi) = 1$.

Concerning the check of falsity of an aggregate, we can start from a 3-DNF formula $\Phi$ and construct an aggregate literal $A' = \#\texttt{sum}\{S\} < 1$, where $S$ is obtained as described above. Then $\Phi$ is a tautology if and only if $A'$ is false with respect to $I = \{\top\}$.

(b) Let $I$ be a partial interpretation for an $\text{LP}_{m,a}^{\mathcal{A}}$ program $\mathcal{P}$ and $A$ an aggregate literal occurring in $\mathcal{P}$. We want to show that deciding whether $A$ is true (resp. false) with respect to $I$ can be done in polynomial-time in the size of $B_{\mathcal{P}}$. By Lemma 18, it is enough to evaluate the aggregate with respect to either $I_{min} = I \cup \neg.(B_{\mathcal{P}} \setminus I)$ or $I_{max} = I \cup (B_{\mathcal{P}} \setminus \neg.I)$. We then end the proof by observing that the interpretations $I_{min}$ and $I_{max}$ can be constructed in polynomial time, and that the value of the aggregate function in $A$ can be computed in polynomial time with respect to the size of the input multiset by assumption. $\square$

In order to prove the tractability of the well-founded semantics we need an efficient method for computing the greatest unfounded set, which is part of the well-founded operator $\mathcal{W}_{\mathcal{P}}$. Hence, we next give a polynomial-time construction of the set $B_{\mathcal{P}} \setminus GUS_{\mathcal{P}}(I)$ by means of a monotone operator.

**Definition 7** Let $I$ be an interpretation for an $\text{LP}_{m,a}^{\mathcal{A}}$ program $\mathcal{P}$. The operator $\phi_I : 2^{B_{\mathcal{P}}} \to 2^{B_{\mathcal{P}}}$ is defined as follows:

$\phi_I(Y) = \{\ell \in B_{\mathcal{P}} \mid \exists \; r \in \mathcal{P} \text{ with } H(r) = \ell \text{ such that}$
$\qquad\qquad$ no (antimonotone) literal in $B^a(r)$ is false with respect to $I$, and
$\qquad\qquad$ all the (monotone) literals in $B^m(r)$ are true with respect to $Y \setminus \neg.I^-\}$

The least fixpoint of $\phi_I$ coincides with the greatest unfounded set of $\mathcal{P}$ with respect to $I$.

**Theorem 20** Let $\mathcal{P}$ be an $\text{LP}_{m,a}^{\mathcal{A}}$ program and $I$ an interpretation for $\mathcal{P}$. Then:

1. The $\phi_I$ operator has a least fixpoint $\phi_I^\omega(\emptyset)$;

2. $GUS_{\mathcal{P}}(I) = B_{\mathcal{P}} \setminus \phi_I^\omega(\emptyset)$.

**Proof.** The $\phi_I$ operator is a monotonically increasing operator in the meet semilattice $\langle B_{\mathcal{P}}, \subseteq \rangle$, and it therefore admits a least fixpoint $\phi_I^\omega(\emptyset)$ (Tarski, 1955). We next prove that $GUS_{\mathcal{P}}(I) = B_{\mathcal{P}} \setminus \phi_I^\omega(\emptyset)$ in two steps:

($\subseteq$) We first observe that $\phi_I^\omega(\emptyset)$ can be computed iteratively, starting from the empty set, as the limit of the sequence $F_0 = \emptyset$, $F_{i+1} = \phi_I(F_i)$. Thus, we prove by induction on $i$ that $GUS_{\mathcal{P}}(I) \subseteq B_{\mathcal{P}} \setminus F_i$ holds. The base case is trivial, since $F_0 = \emptyset$ by definition and $GUS_{\mathcal{P}}(I)$ is a subset of $B_{\mathcal{P}}$ by Definition 1. We then assume $GUS_{\mathcal{P}}(I) \subseteq B_{\mathcal{P}} \setminus F_i$ in order to prove that $GUS_{\mathcal{P}}(I) \subseteq B_{\mathcal{P}} \setminus F_{i+1}$. Since $GUS_{\mathcal{P}}(I)$ is an unfounded set for $\mathcal{P}$ with respect to $I$ by Theorem 2, by Definition 1 we have that, for each $\ell \in GUS_{\mathcal{P}}(I)$ and for each rule $r \in \mathcal{P}$ with $H(r) = \ell$, either (1) some (antimonotone) literal in $B^a(r)$ is false with respect to $I$, or (2) some (monotone) literal in $B^m(r)$ is false with respect to $I \mathbin{\dot{\cup}} \neg.GUS_{\mathcal{P}}(I)$. We want to show that such a $\ell$ does not belong to $F_{i+1}$, that is, each rule $r$ as above is such that either (i) some (antimonotone) literal in $B^a(r)$ is false with respect to $I$, or (ii) some (monotone) literal in $B^m(r)$ is not true with respect to $F_i \setminus \neg.I^-$ (recall that $F_{i+1} = \phi_I(F_i)$ by definition). Since (1) and (i) are equals, we have to show that (2) implies (ii). To this end, assume that there is a (monotone) literal $\ell' \in B^m(r)$ which is false with respect to $I \mathbin{\dot{\cup}} \neg.GUS_{\mathcal{P}}(I)$. Our aim is to show that $\ell'$ is false with respect to $J = (F_i \setminus \neg.I^-) \cup \neg.(B_{\mathcal{P}} \setminus (F_i \setminus \neg.I^-))$, since in this case $\ell'$ would be not true with respect to $F_i \setminus \neg.I^-$ (see Lemma 18). We start by proving that $(I \mathbin{\dot{\cup}} \neg.GUS_{\mathcal{P}}(I))^- = I^- \cup \neg.GUS_{\mathcal{P}}(I)$ is a subset of $J^-$. Observe that $J^- = \neg.(B_{\mathcal{P}} \setminus (F_i \setminus \neg.I^-)) = I^- \cup \neg.(B_{\mathcal{P}} \setminus F_i)$ because $\neg.I^-$ is a subset of $B_{\mathcal{P}}$. Thus, since $GUS_{\mathcal{P}}(I) \subseteq B_{\mathcal{P}} \setminus F_i$ by induction hypothesis, we obtain $(I \mathbin{\dot{\cup}} \neg.GUS_{\mathcal{P}}(I))^- = I^- \cup \neg.GUS_{\mathcal{P}}(I) \subseteq I^- \cup \neg.(B_{\mathcal{P}} \setminus F_i) = J^-$. Since $J$ is total, $(I \mathbin{\dot{\cup}} \neg.GUS_{\mathcal{P}}(I))^- \subseteq J^-$ implies that there is an extension $K$ of $I \mathbin{\dot{\cup}} \neg.GUS_{\mathcal{P}}(I)$ such that $K^- \subseteq J^-$ and $K^+ \supseteq J^-$ (for example, the one containing as true all the standard positive literals which are undefined with respect to $I \mathbin{\dot{\cup}} \neg.GUS_{\mathcal{P}}(I)$). Since $\ell'$ is false with respect to $I \mathbin{\dot{\cup}} \neg.GUS_{\mathcal{P}}(I)$ by assumption and $K$ is an extension of $I \mathbin{\dot{\cup}} \neg.GUS_{\mathcal{P}}(I)$, $\ell'$ is false with respect to $K$ by Remark 1. Thus, since $J \leq K$ and $\ell'$ is monotone, the latter implies that $\ell'$ is false with respect to $J$ as well.

($\supseteq$) We prove that $B_{\mathcal{P}} \setminus \phi_I^\omega(\emptyset)$ is an unfounded set for $\mathcal{P}$ with respect to $I$, that is, for each $r \in \mathcal{P}$ with $H(r) \in B_{\mathcal{P}} \setminus \phi_I^\omega(\emptyset)$, either (1) some (antimonotone) literal in $B^a(r)$ is false with respect to $I$, or (2) some (monotone) literal in $B^m(r)$ is false with respect to $I \mathbin{\dot{\cup}} \neg.(B_{\mathcal{P}} \setminus \phi_I^\omega(\emptyset))$. By Definition 7, $H(r) \notin \phi_I^\omega(\emptyset)$ implies either that (i) some (antimonotone) literal in $B^a(r)$ is false with respect to $I$, or that (ii) some (monotone) literal in $B^m(r)$ is not true with respect to $\phi_I^\omega(\emptyset) \setminus \neg.I^-$. Since (i) and (1) are equals, we have to show that (ii) implies (2). To this end, assume that there is a (monotone) literal $\ell \in B^m(r)$ which is not true with respect to $\phi_I^\omega(\emptyset) \setminus \neg.I^-$. Thus, there is an extension of $\phi_I^\omega(\emptyset) \setminus \neg.I^-$ for which $\ell$ is false, and in particular $\ell$ must be false with respect to $J = (\phi_I^\omega(\emptyset) \setminus \neg.I^-) \cup \neg.(B_{\mathcal{P}} \setminus (\phi_I^\omega(\emptyset) \setminus \neg.I^-))$ because of Lemma 18. Now observe that $(I \mathbin{\dot{\cup}} \neg.(B_{\mathcal{P}} \setminus \phi_I^\omega(\emptyset)))^- = I^- \cup \neg.(B_{\mathcal{P}} \setminus \phi_I^\omega(\emptyset)) = \neg.(B_{\mathcal{P}} \setminus (\phi_I^\omega(\emptyset) \setminus \neg.I^-)) =$ $J^-$ holds (because $\neg.I^- \subseteq B_{\mathcal{P}}$), and so $(I \,\dot\cup\, \neg.(B_{\mathcal{P}} \setminus \phi_I^\omega(\emptyset)))^+ \subseteq J^+$ because $J$ is total. By combining the last two sentences we obtain $I \,\dot\cup\, \neg.(B_{\mathcal{P}} \setminus \phi_I^\omega(\emptyset)) \le J$. Therefore, since $\ell$ is a monotone literal which is false with respect to $J$, the latter implies that $\ell$ is false with respect to $I \,\dot\cup\, \neg.(B_{\mathcal{P}} \setminus \phi_I^\omega(\emptyset))$ as well, and so (2) holds. □

Eventually, Property 5 is a consequence of the following theorem. As mentioned earlier, this theorem also follows from Theorem 7.4 in the work of Pelov et al. (2007) because of Corollary 25, but the proof provided here differs considerably from the one of Theorem 7.4 in the work of Pelov et al.

**Theorem 21** Given an $\mathrm{LP}_{m,a}^{\mathcal{A}}$ program $\mathcal{P}$:

1. The greatest unfounded set $GUS_{\mathcal{P}}(I)$ of $\mathcal{P}$ with respect to a given interpretation $I$ is polynomial-time computable;

2. $\mathcal{W}_{\mathcal{P}}^\omega(\emptyset)$ is polynomial-time computable.

**Proof.** (1.) From Theorem 20, $GUS_{\mathcal{P}}(I) = B_{\mathcal{P}} \setminus \phi_I^\omega(\emptyset)$. We next show that $\phi_I^\omega(\emptyset)$ is efficiently computable. The fixpoint $\phi_I^\omega(\emptyset)$ is the limit $\phi_\lambda$ of the sequence $\phi_0 = \emptyset$, $\phi_k = \phi_I(\phi_{k-1})$. This limit is reached in a polynomial number of applications of $\phi_I$ because each new element of the sequence $\phi_k$ must add at least a new atom (otherwise the limit has been already reached), that is, $\lambda \le |B_{\mathcal{P}}|$. If we show that each application of $\phi_I$ is feasible in polynomial time, we can conclude that $\phi_\lambda$ is computable in polynomial time. Each step processes at most all the rules once, and for each rule checks the truth-value of at most all body literals once. The check of the truth valuation is clearly tractable for all standard (i.e., non-aggregates) literals; the tractability of the check for aggregate literals stems from Proposition 19, as we deal with monotone and antimonotone aggregate atoms only. In conclusion, $\phi_\lambda$ is computable in polynomial time, and $GUS_{\mathcal{P}}(I)$ is tractable as well since it is obtainable as $B_{\mathcal{P}} \setminus \phi_I^\omega(\emptyset)$.

(2.) By the argumentation carried out for $\phi_I^\omega(\emptyset)$, we can show that $\mathcal{W}_{\mathcal{P}}^\omega(\emptyset)$ is computed in a number of steps which is polynomial (actually linear) in $|B_{\mathcal{P}}|$. Indeed, each step is polynomial-time computable: We have just proved the tractability of $GUS_{\mathcal{P}}(I)$, and $\mathcal{T}_{\mathcal{P}}$ is polynomial-time computable as well. □

This result has a positive impact also for the computation of the answer set semantics of logic programs with aggregates. Indeed, as stated in Theorem 16, $\mathcal{W}_{\mathcal{P}}^\omega(\emptyset)$ approximates the intersection of all answer sets (if any) from the bottom, and can be therefore used to efficiently prune the search space. It is worthwhile noting that the computation of the well-founded semantics is also hard for polynomial-time. In particular, deciding whether a (ground) atom is true with respect to the well-founded semantics is P-complete, as this task is P-hard even for the standard well-founded semantics of aggregate-free programs (and, from Proposition 5, our semantics coincides with the standard well-founded on aggregate-free programs).

We end this section by briefly addressing the complexity of non-ground programs. When considering data-complexity (i.e., an $\mathrm{LP}_{m,a}^{\mathcal{A}}$ program $\mathcal{P}$ is fixed and the input only consists of facts), the results are as for propositional programs: Deciding whether a (ground) atom is true with respect to the well-founded semantics of a non-ground program is P-complete, under data-complexity (Van Gelder et al., 1991). However, if program complexity (i.e., an $LP_{m,a}^{\mathcal{A}}$ program $\mathcal{P}$ is given as input) is considered, complexity of reasoning rises exponentially. Indeed, a non-ground program $\mathcal{P}$ can be reduced, by naive instantiation, to a ground instance of the problem, and in general the size of $Ground(\mathcal{P})$ is single exponential in the size of $\mathcal{P}$. The complexity of reasoning increases accordingly by one exponential, from P to EXPTIME, and the result can be derived using complexity upgrading techniques (Eiter, Gottlob, & Mannila, 1997; Gottlob, Leone, & Veith, 1999).

## 6. Compilation into Standard LP, Implementation and Experimental Results

The well-founded semantics for $LP_{m,a}^{\mathcal{A}}$ programs has been implemented by extending the DLV system (Leone et al., 2006). In this section we briefly describe the implemented prototype and report on the results of our experiments aimed at assessing its efficiency. Note that, even if $LP_{m,a}^{\mathcal{A}}$ programs can be replaced by equivalent LP programs (for a rewriting strategy see Section 6.1 below), our experimental results highlight a significant performance advantage of $LP_{m,a}^{\mathcal{A}}$ encodings.

### 6.1 Compilation into Standard Logic Programming

In this section we briefly present a strategy for representing #count, #sum and #times with standard constructs.[2] The compilation is in the spirit of the one introduced for #min and #max by Alviano, Faber, and Leone (2008) and defines a subprogram computing the value of a (possibly recursive) aggregate. The compilation takes into account specific properties of monotone and antimonotone aggregate functions, and is therefore referred to as *monotone/antimonotone encoding* (*mae*).

The monotone/antimonotone encoding of an $LP_{m,a}^{\mathcal{A}}$ program $\mathcal{P}$ is obtained by replacing each aggregate literal $A = f(S) \prec T$ by a new predicate symbol $f_{\prec}$. Predicate $f_{\prec}$ is defined by means of a subprogram (i.e., a set of rules) that can be thought of as a compilation of $A$ into standard LP. The compilation uses a total order $<$ of the elements of $U_{\mathcal{P}} \cup \{\perp\}$, where $\perp$ is a symbol not occurring in $\mathcal{P}$ and such that $\perp < u$ for each $u \in U_{\mathcal{P}}$. We further assume the presence of a "built-in" relation $\overline{Y} < \overline{Y'}$, where $\overline{Y} = Y_1, \ldots, Y_n$ and $\overline{Y'} = Y_1', \ldots, Y_n'$ are lists of terms. This built-in relation has $\overline{y} < \overline{y'}$ if and only if $\overline{y}$ precedes $\overline{y'}$ in the lexicographical order induced by $<$. Moreover, we will use a built-in relation $\overline{Y} \le \overline{Y'}$, where $\overline{y} \le \overline{y'}$ is true if and only if either $\overline{y} < \overline{y'}$ or $\overline{y} = \overline{y'}$. For simplicity, let us assume that $A$ is of the form $f(\{\overline{Y} : p(\overline{Y}, \overline{Z})\}) \prec k$, where $\overline{Y}$ and $\overline{Z}$ are lists of local variables and $k$ is an integer constant. For such an aggregate, we introduce a new predicate symbol $f_{aux}$ of arity $|\overline{Y}| + 1$ and rules for modeling that an atom $f_{aux}(\overline{y}, s)$ must be true whenever the value of $f(\{\overline{Y} : p(\overline{Y}, \overline{Z}), \ \overline{Y} \le \overline{y}\})$ is at least $s$. Thus, we use a fact for representing the value of the aggregate function for the empty set, and a rule for increasing this value for larger sets. The lexicographical order induced by $<$ is used to guarantee that all elements in the set are

---

2. Since we are considering only monotone and antimonotone aggregate literals, the domains of #sum and #times are assumed to be $\mathbb{N}$ and $\mathbb{N}^+$, respectively.

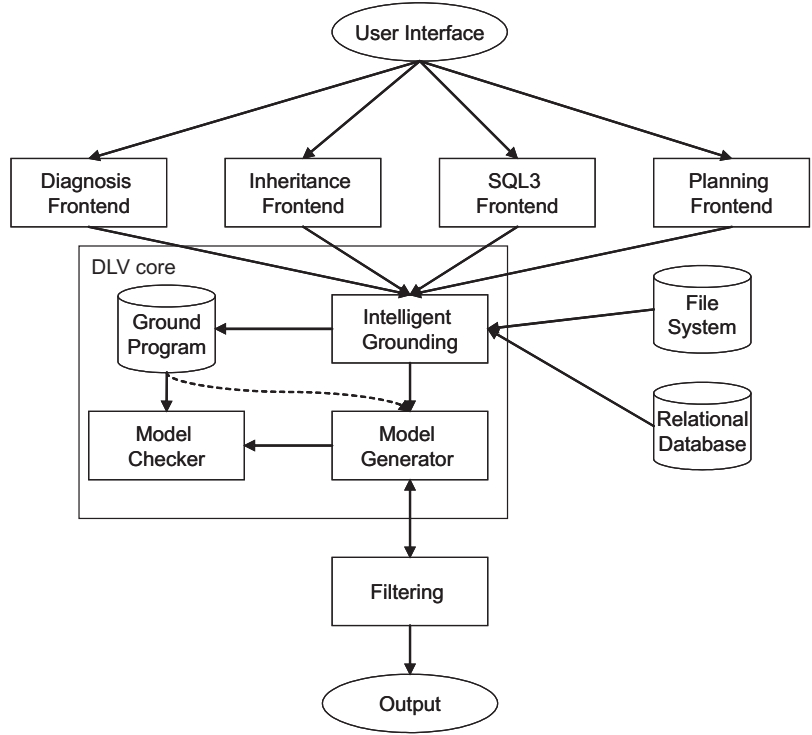

Figure 2: Prototype system architecture.

considered at most once. In particular, the following rules are introduced:

$$f_{aux}(\overline{\perp}, \alpha).$$
$$f_{aux}(\overline{Y'}, X) := f_{aux}(\overline{Y}, S),\ p(\overline{Y'}, \overline{Z}), \qquad \text{where} \quad \begin{cases} \alpha = 0,\ \beta = S+1 & \text{if } f = \#\texttt{count}; \\ \alpha = 0,\ \beta = S+Y'_1 & \text{if } f = \#\texttt{sum}; \\ \alpha = 1,\ \beta = S \times Y'_1 & \text{if } f = \#\texttt{times}. \end{cases}$$
$$\overline{Y} < \overline{Y'},\ X = \beta.$$

If $\prec \in \{\geq, >\}$, truth of an aggregate $f(\{\overline{Y} : p(\overline{Y}, \overline{Z})\}) \prec k$ must be inferred if and only if *some* atom $f_{aux}(\overline{y}, s)$ such that $s \prec k$ is true. This aspect is modeled by means of the following rules:

$$f_{\geq k} := f_{aux}(\overline{Y}, S),\ S \geq k. \qquad\qquad f_{>k} := f_{aux}(\overline{Y}, S),\ S > k.$$

If $\prec$ is $\leq$, instead, truth of an aggregate $f(\{\overline{Y} : p(\overline{Y}, \overline{Z})\}) \leq k$ must be inferred if and only if *all* atoms $f_{aux}(\overline{y}, s)$ such that $s > k$ are false (and similar if $\prec$ is $<$). These aspects are modeled by means of the following rules:

$$f_{\leq k} := \texttt{not}\ f_{>k}. \qquad\qquad f_{<k} := \texttt{not}\ f_{\geq k}.$$

Extending the technique to aggregate literals with global variables is quite simple: Global variables are added to the arguments of all the atoms used in the compilation, and a new predicate $f_{group-by}$ is used for collecting their possible substitutions.

## 6.2 System Architecture and Usage

We have extended DLV by implementing the well-founded operator and the well-founded semantics for $\mathrm{LP}_{m,a}^{\mathcal{A}}$ programs described in this paper. The architecture of the prototype is reported in Figure 2. In detail, we modified two modules of DLV, the *Intelligent Grounding* module and the *Model Generator* module. In our prototype, the well-founded semantics is adopted if one of `-wf` or `--well-founded` is specified on the command-line. Otherwise, the stable model semantics is adopted as usual. The well-founded operator $\mathcal{W}_{\mathcal{P}}$ introduced in Section 3 is used for both semantics. In particular, for the stable model semantics, the well-founded model is profitably used for pruning the search space. For the well-founded semantics, the well-founded model is printed after the computation of the least fixpoint of the well-founded operator. In this case the output of the system consists of two sets, for representing true and undefined standard atoms in the well-founded model. A binary of the prototype is available at `http://www.dlvsystem.com/dlvRecAggr/`.

## 6.3 Experimental Results

To our knowledge, the implemented prototype is currently the only system supporting a well-founded semantics for logic programs with recursive aggregates. For certain special cases, such as when the well-founded model is total, the well-founded model coincides with other semantics such as answer sets (see Corollary 17) and in theses cases systems supporting those semantics such as IDP (Wittocx, Mariën, & Denecker, 2008), Smodels (Simons et al., 2002), or clasp (Gebser, Kaufmann, Neumann, & Schaub, 2007), can be used to compute the well-founded model.

We are however interested in systems that are able to compute the well-founded model for all input programs. One of the major systems supporting the well-founded semantics, XSB (Swift & Warren, 2010), has some support for aggregates, but (apart from #min and #max) XSB does not support *recursive* aggregates (i.e., aggregates occurring in recursive definitions). Therefore, our experiments have been designed for investigating the computational behavior of aggregate constructs with respect to equivalent encodings without aggregates.

More specifically, we introduce the *Attacks* problem, which is inspired by the classic *Win-Lose* problem often used in the context of the well-founded semantics for standard logic programs, and study performance on it.

**Definition 8 (Attacks Problem)** In the Attacks problem, a set of $p$ players and a positive integer $m$ are given. Each player attacks $n$ other players. A player wins if no more than $m$ winners attack it. This kind of problem is frequently present in turn-based strategy games.

Note that the definition of winner is recursive and, in particular, a recursive aggregate is the natural way of encoding this problem.

**Example 14** An instance of the Attacks problem in which $p = 6$, $n = 2$ and $m = 1$ could be the following:

- player $a$ attacks players $b$ and $c$;
- player $b$ attacks players $a$ and $c$;
- player $c$ attacks players $a$ and $b$;
- player $d$ attacks players $b$ and $f$;
- player $e$ attacks players $c$ and $f$;
- player $f$ attacks players $d$ and $e$.

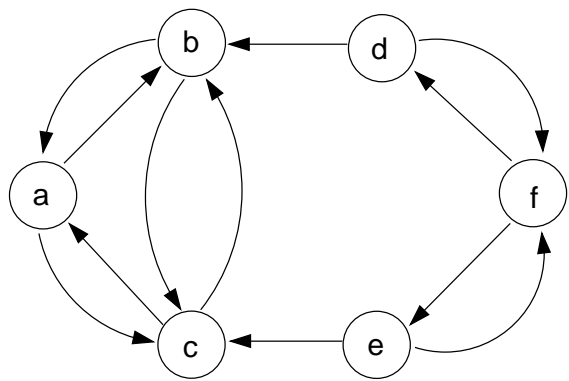

Figure 3: An instance of the *Attacks* problem with 6 players, each one attacking 2 other players.

A graphical representation of this instance is shown in Figure 3. Since $d$ is only attacked by $f$, we can conclude that $d$ is a winner. Similarly for $e$. Therefore, $f$ is not a winner because $f$ is attacked by $d$ and $e$, which are winners. For the other players, namely $a$, $b$ and $c$, we cannot determine who is a winner or not.

In our experiments, instances of Attacks are encoded by means of the predicates $max$, $player$ and $attacks$ for representing the parameter $m$, the set of players and the attacks of the players, respectively. We consider three equivalent encodings for the Attacks problem.

### 6.3.1 Aggregate-Based Encoding

This encoding is a natural representation of the Attacks problem in $LP^{\mathcal{A}}_{m,a}$. The complete encoding consists of a single rule, reported below:

$$win(X) :\!- max(M),\ player(X),\ \#\texttt{count}\{Y : attacks(Y,X),\ win(Y)\} \le M.$$

### 6.3.2 Join-Based Encoding

An equivalent encoding can be obtained by computing a number of joins proportional to $m$. The tested encoding is reported below:

$$\begin{aligned}
&win(X) :\!- player(X),\ \texttt{not}\ lose(X).\\
&lose(X) :\!- max(1),\ attacks(Y_1,X),\ win(Y_1),\\
&\qquad\qquad\qquad attacks(Y_2,X),\ win(Y_2),\ Y_1 < Y_2.\\
&lose(X) :\!- max(2),\ attacks(Y_1,X),\ win(Y_1),\\
&\qquad\qquad\qquad attacks(Y_2,X),\ win(Y_2),\ Y_1 < Y_2,\\
&\qquad\qquad\qquad attacks(Y_3,X),\ win(Y_3),\ Y_1 < Y_3,\ Y_2 < Y_3.\\
&lose(X) :\!- max(3),\ \dots
\end{aligned}$$

Note that in the encoding above there is a rule for each possible value of parameter $m$. However, only one of these rules is considered by our solver during program instantiation. In fact, only the rule is instantiated, which contains the instance of atom $max(m)$ for which a fact is present. All the other rules are satisfied because of a false body literal.

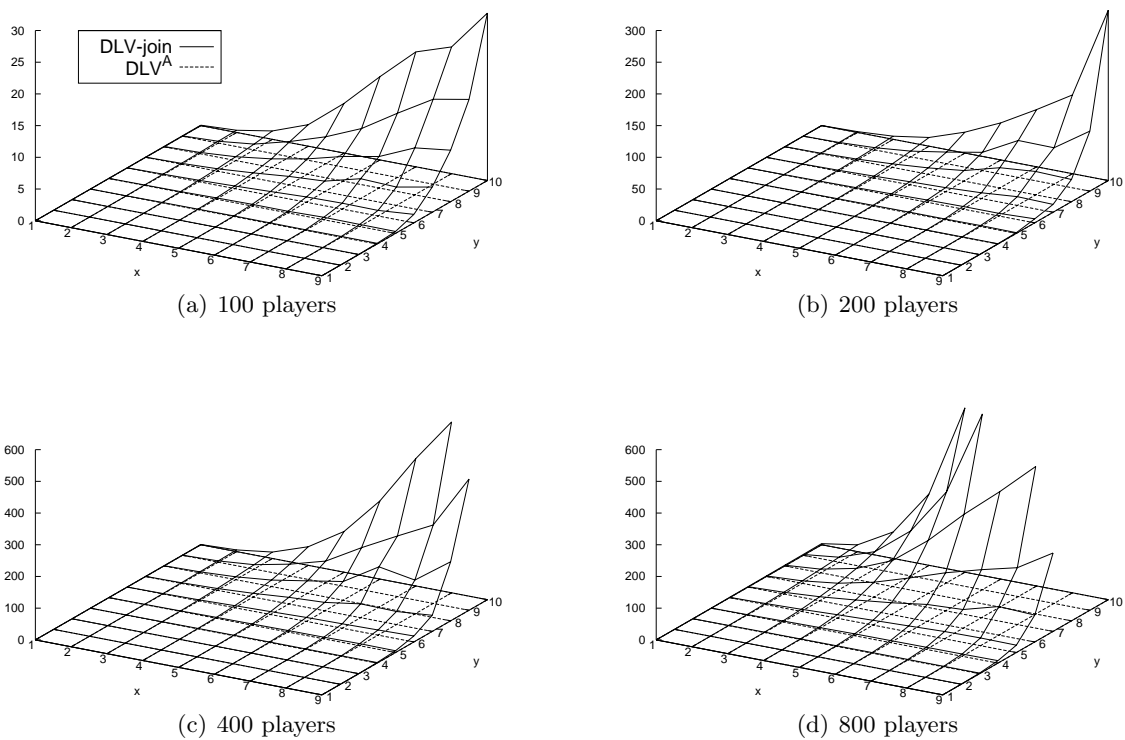

Figure 4: *Attacks:* Average execution time of DLV running the aggregate-based encoding and DLV running the join-based encoding.

### 6.3.3 Mae-Based Encoding

This encoding has been obtained by applying the compilation presented in Section 6.1 with some minor simplifications. The full encoding is reported below:

$$win(X) :\!\!- player(X), \texttt{not}\ lose(X).$$
$$lose(X) :\!\!- count(X, Y, S),\ max(M),\ S > M.$$
$$count(X, Y, 1) :\!\!- aux(X, Y).$$
$$count(X, Y', S') :\!\!- count(X, Y, S),\ aux(X, Y'),\ Y < Y',\ S' = S + 1.$$
$$aux(X, Y) :\!\!- attacks(Y, X),\ win(Y).$$

Intuitively, an atom $count(x, y, s)$ stands for "there are at least $s$ constants $y'$ such that $y' \leq y$ and $attacks(y', x)$, $win(y')$ is true". Note that the rules defining predicate $count$ use the natural order of integers to guarantee that each $y'$ is counted at most once.

**Example 15** The instance shown in Figure 3 is represented by means of the following facts:

$player(a).$     $player(b).$     $player(c).$     $player(d).$     $player(e).$     $player(f).$
$attacks(a, b).$   $attacks(b, a).$   $attacks(c, a).$   $attacks(d, b).$   $attacks(e, c).$   $attacks(f, d).$
$attacks(a, c).$   $attacks(b, c).$   $attacks(c, b).$   $attacks(d, f).$   $attacks(e, f).$   $attacks(f, e).$
$max(1).$

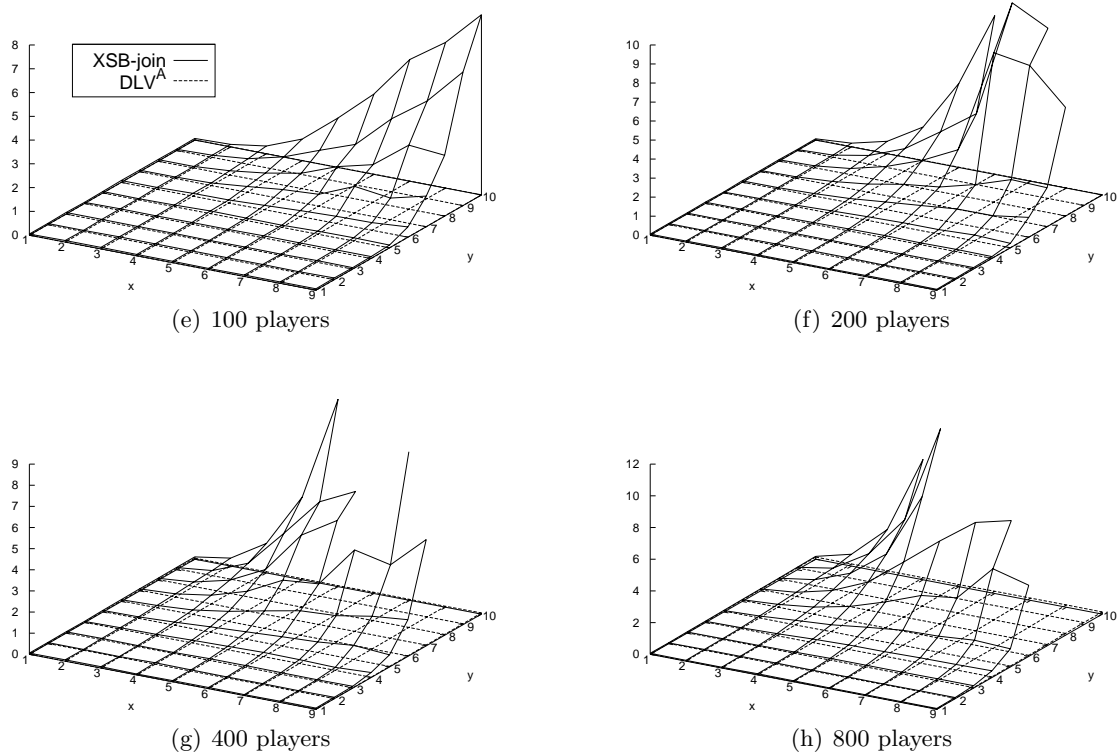

Figure 5: *Attacks:* Average execution time of DLV running the aggregate-based encoding and XSB running the join-based encoding.

For all the encodings, the well-founded model restricted to the *win* predicate is $\{win(d)$, $win(e)$, `not` $win(f)\}$. Note that $win(a)$, $win(b)$ and $win(c)$ are neither true nor false, and so they are undefined.

### 6.3.4 DISCUSSION

We performed an intensive experimentation for this benchmark by varying the parameters $p$, $m$ and $n$. For each combination of these parameters, we measured the average execution time of DLV and XSB (version 3.2) on 3 randomly generated instances. The experiments have been performed on a 3GHz Intel® Xeon® processor system with 4GB RAM under the Debian 4.0 operating system with GNU/Linux 2.6.23 kernel. The DLV prototype used has been compiled with GCC 4.4.1. For every instance, we have allowed a maximum running time of 600 seconds (10 minutes) and a maximum memory usage of 3GB.

The results of our experimentation are reported in Figures 4–7. In the graphs, DLV[A] is the implemented prototype with the aggregate-based encoding, DLV-join and DLV-mae the implemented prototype with the aggregate-free encodings, XSB-join and XSB-mae the XSB system with the aggregate-free encodings (as mentioned earlier, XSB does not support recursive aggregates). For the XSB system, we explicitly set indices and tabled predicates for optimizing its computation.

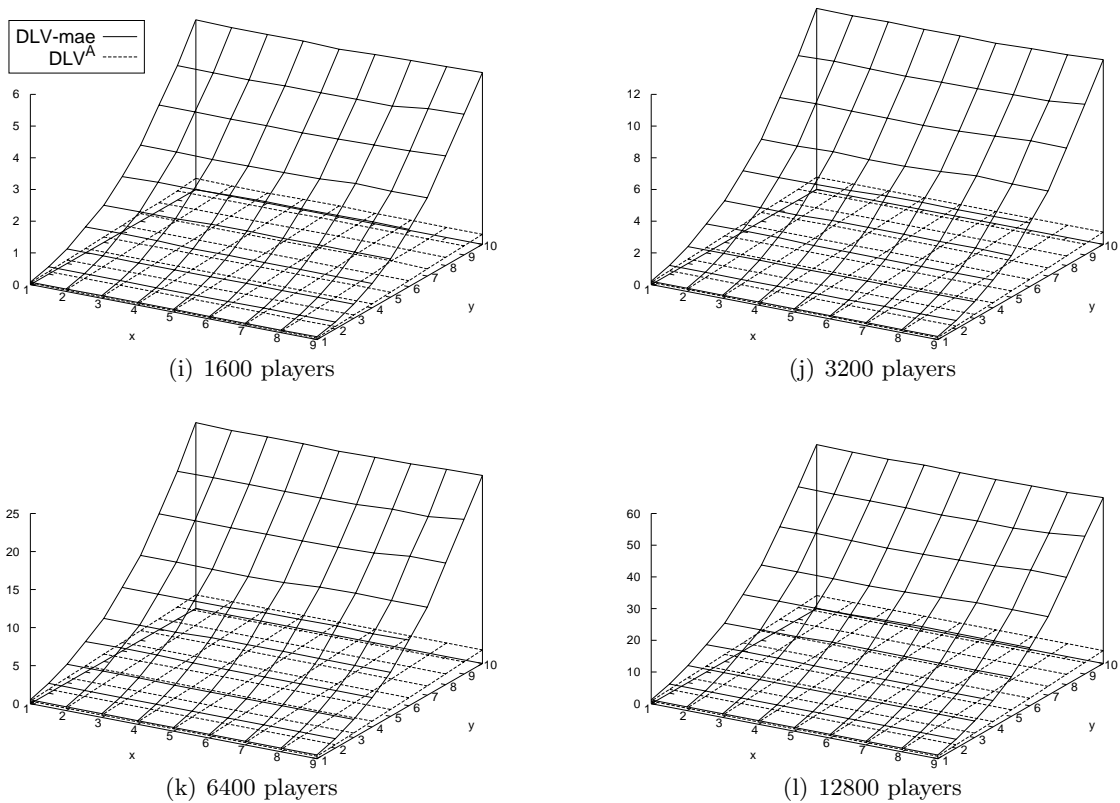

Figure 6: *Attacks:* Average execution time of DLV running the aggregate-based encoding and DLV running the *mae*-based encoding.

For each graph, the number of players is fixed, while parameters $m$ (x-axis) and $n$ (y-axis) vary. Therefore, the size of the instances grows moving from left to right along the y-axis, while it is invariant with respect to the x-axis. However, the number of joins required by the join-based encoding depends on the parameter $m$. As a matter of fact, we can observe in the graphs in Figures 4–5 that the average execution time of the join-based encoding increases along both the x- and y-axis (for both DLV and XSB). Instead, for the encoding using aggregates, and for the mae-based encoding, the average execution time only depends on instance sizes, as shown in the graphs in Figures 6–7.

For the join-based encoding, XSB is generally faster than DLV, but consumes much more memory. Indeed, in Figure 5, we can observe that XSB terminates its computation in a few seconds for the smallest instances, but rapidly runs out of memory on slightly larger instances. Considering the mae-based encoding, we can observe significant performance gains for both DLV and XSB (see Figures 6–7). Indeed, both systems complete their computation in the allowed time and memory on larger instances. Computational advantages of the mae-based encoding with respect to the join-based encoding are particularly evident for XSB, which solved all tested instances with this encoding. However, also XSB with the mae-based encoding is outperformed by DLV with native support for aggregate constructs (see Figure 7).

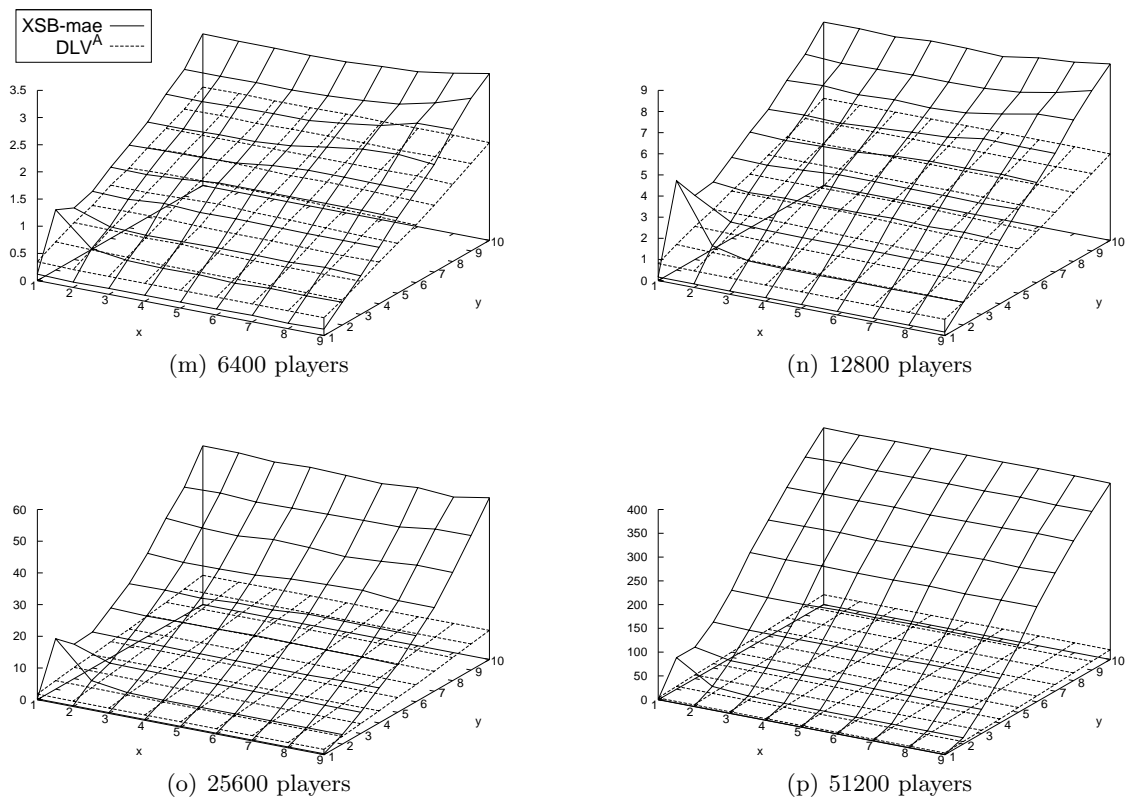

Figure 7: *Attacks:* Average execution time of DLV running the aggregate-based encoding and XSB running the *mae*-based encoding.

In sum, the experimental results highlight that the presence of aggregate constructs can significantly speed-up the computation. Indeed, the encoding using recursive aggregates outperforms the aggregate-free encodings in all tested instances.

# 7. Related Work

Defining a well-founded semantics for logic programs with aggregates has been a challenge of major interest in the last years. The first attempts, not relying on a notion of unfounded set, have been defined on a restricted language. Some of these are discussed by Kemp and Stuckey (1991). Another semantics falling in this class is the one introduced by Van Gelder (1992), subsequently generalized by Osorio and Jayaraman (1999). The main problem of these semantics is that they often leave too many undefined literals, as shown by Ross and Sagiv (1997).

A first attempt to define a well-founded semantics for unrestricted $LP^{\mathcal{A}}$ has been done by Kemp and Stuckey (1991). This semantics is based on a notion of unfounded sets. According to Kemp and Stuckey, a set $\mathcal{X}$ of standard atoms is an unfounded set for a (ground) program $\mathcal{P}$ with respect to an interpretation $I$ if, for each rule $r \in \mathcal{P}$ with $H(r) \in \mathcal{X}$, either (a) some literal in $B(r)$ is false with respect to $I$, or (b) $B(r) \cap \mathcal{X} \neq \emptyset$. Note that only standard literals are considered by condition (b), and aggregates are not covered by it. We point out that this definition of unfounded set makes the semantics inadequate for programs with recursive aggregates, even if only monotone aggregates are considered. For example, for the program $\{a(1){:}{-}\#\mathtt{count}\{X : a(X)\} > 0.\}$, the well-founded model in the work of Kemp and Stuckey is $\emptyset$, while a reasonable well-founded semantics should identify $a(1)$ as false.

Pelov et al. (2007) defined a well-founded semantics based on approximating operators, namely $\tilde{\mathcal{D}}$-well-founded semantics, which extends the standard well-founded semantics; indeed, they coincide for aggregate-free programs. More in detail, in that work aggregates are evaluated in one of three possible ways. Therefore, a family of semantics is defined by Pelov et al., which can be ordered by precision: More precise three-valued aggregates lead to more precise semantics. In general, higher precision comes at the price of a higher computational complexity. The authors discuss the following three-valued aggregate relations for the evaluation of aggregate literals: *trivial*, *bound* and *ultimate approximating aggregates*, where the first is the less precise, and the last is the most precise. Semantics relying on trivial approximating aggregates is very imprecise, but it is still suitable for the class of stratified aggregate programs. Both trivial and bound approximations have polynomial complexity, while ultimate has been shown to be intractable for nonmonotone aggregate functions (Pelov, 2004). A detailed comparison with our results is presented in Section 7.1.

Ferraris (2005) showed that the semantics of SMODELS programs with positive weight constraints is equal to answer sets as defined by Faber et al. (2004) on the respective fragment. Since by Theorem 16 $\mathcal{W}_{\mathcal{P}}^{\omega}(\emptyset)$ approximates answer sets as defined by Faber et al., $\mathcal{W}_{\mathcal{P}}^{\omega}(\emptyset)$ can be used also as an approximating operator for the respective SMODELS programs. Indeed, it can be shown that the *AtMost* pruning operator of SMODELS (Simons et al., 2002) is a special case of the $\phi_I$ operator (defined in the proof of Theorem 21).

Other works attempted to define stronger notions of well-founded semantics (also for programs with aggregates), like the Ultimate Well-Founded Semantics (Denecker et al., 2001), or WFS[1] and WFS[2] (Dix & Osorio, 1997). Whether a characterization of these semantics in terms of unfounded sets can exist for these semantics is unclear and left for future research.

Concerning compilations of $LP^{\mathcal{A}}$ programs into standard LP, a transformation was provided by Van Gelder (1992). The compilation that we presented in Section 6.1 differs from the one introduced by Van Gelder in several respects. Our approach uses a total order of the universe of the input program and takes advantage of the character of monotonicity/antimonotonicity of the aggregate literals in the input program, while the transformation defined by Van Gelder uses uninterpreted function symbols for representing ground sets, and recursive negation for checking truth of aggregate literals. We briefly discuss these aspects in the following. Roughly, for an aggregate $f(S) \prec k$, uninterpreted function symbols are used by the transformation in the work of Van Gelder for determining all pairs $S', k'$ such that $S'$ is a ground set associated with $S$ and $k' = f(S')$. After that, the transformation defined by Van Gelder checks whether there exists a pair $S', k'$ satisfying the following conditions: (i) for every element $\langle consts : conj \rangle$ in $S'$, $conj$ is true; (ii) $k' \prec k$ holds. We point out that Condition (i) requires recursive negation in order to be checked. Indeed, it is equivalent to "there is no element $\langle consts : conj \rangle$ in $S'$ such that $conj$ is not true." This aspect of the transformation has an undesirable side effect: Stratified $\text{LP}_{m,a}^{\mathcal{A}}$ programs may have partial well-founded models, that is, Theorem 9 does not hold for programs compiled with the transformation introduced by Van Gelder. An example of this side effect is given by Van Gelder, where it is shown that this transformation possibly leads to partial well-founded models for instances of *Company Controls*, a well-known problem that can be modeled by using monotone recursive aggregates.

## 7.1 Comparison with the work of Pelov et al. (2007)

In this section we report a detailed comparison of the well-founded semantics as defined in this paper with the one of Pelov et al. (2007). We recall that Pelov et al. defines well-founded and stable semantics as the least and total fixpoints of the three-valued stable model operator extended to aggregate programs.

We start by observing that the evaluation of *ultimate approximating aggregates* coincides with the evaluation of aggregates defined in this article; also the evaluation of *bound approximating aggregates* coincides for monotone and antimonotone aggregates (as a consequence of Lemma 18 in this paper and Proposition 7.16 in the work of Pelov et al., 2007).

Let us now introduce a translation of an aggregate literal into a formula of standard literals. For a (partial) interpretation $I$, let $conj(I)$ denote the conjunction of all the literals in $I$. The translation $trm(A)$ of a ground aggregate literal $A$ is defined as follows:

$$trm(A) = \bigvee \{ conj(I) \mid \ I \text{ is a subset-minimal interpretation}$$
$$\text{such that } A \text{ is true with respect to } I \}$$

Note that, for each (partial) interpretation $J$, the evaluation of $A$ with respect to $J$ coincides with the evaluation of $trm(A)$ with respect to $J$ (Proposition 2 and Proposition 3 in the work of Pelov et al., 2003). Moreover, for a monotone (resp. antimonotone) aggregate literal $A$, only positive (resp. negative) literals appear in $trm(A)$.

For a rule $r$ in a ground $\text{LP}_{m,a}^{\mathcal{A}}$ program $\mathcal{P}$ and an aggregate literal $A \in B(r)$, the translation $trm(\mathcal{P}, r, A)$ of $A$ in $r$ is the program obtained from $\mathcal{P}$ by removing $r$ and by adding a rule $r'$ such that $H(r') = H(r)$ and $B(r') = B(r) \setminus \{A\} \cup conj$, for each $conj \in trm(A)$. Therefore, the full translation $trm(\mathcal{P})$ of $\mathcal{P}$ is defined as the recursive application of $trm(\mathcal{P}, r, A)$ (note that the order in which rules and aggregates are processed is not relevant). We next show that $\mathcal{P}$ and $trm(\mathcal{P})$ have the same unfounded sets.

**Lemma 22** *A set of atoms $\mathcal{X}$ is an unfounded set for a program $\mathcal{P}$ with respect to an interpretation $I$ if and only if $\mathcal{X}$ is an unfounded set for $trm(\mathcal{P})$ with respect to $I$.*

**Proof.** We use induction on the number of aggregate literals in $\mathcal{P}$. If $\mathcal{P}$ has no aggregate literals, then $\mathcal{P} = trm(\mathcal{P})$. Now consider a program $\mathcal{P}$ and a rule $r \in \mathcal{P}$ with an aggregate literal $A$ in $B(r)$. We want to show that a set $\mathcal{X}$ of atoms is an unfounded set for $\mathcal{P}$ with respect to $I$ if and only if $\mathcal{X}$ is an unfounded set for $trm(\mathcal{P}, r, A)$ with respect to $I$, since in this case we might apply the induction hypothesis and prove the claim. Thus, we can end the proof by means of the following observations: (i) $A$ is false with respect to an interpretation $J$ if and only if $trm(A)$ is false with respect to $J$, that is, if and only if for each conjunction $conj \in trm(A)$ there is a literal $\ell \in conj$ such that $\ell$ is false with respect to $J$; (ii) such an $\ell$ is a positive (resp. negative) standard literal if and only if $A$ is monotone (resp. antimonotone). □

We can then prove that the well-founded operators of $\mathcal{P}$ and $trm(\mathcal{P})$ coincide.

**Lemma 23** *Let $\mathcal{P}$ be an $\mathrm{LP}_{m,a}^{\mathcal{A}}$ program and $I$ an interpretation for $\mathcal{P}$. Then $\mathcal{W}_{\mathcal{P}}(I) = \mathcal{W}_{trm(\mathcal{P})}(I)$.*

**Proof.** We have to show that (1) $\mathcal{T}_{\mathcal{P}}(I) = \mathcal{T}_{trm(\mathcal{P})}(I)$ and (2) $GUS_{\mathcal{P}}(I) = GUS_{trm(\mathcal{P})}(I)$. We note that (2) immediately follows from Lemma 22. In order to prove (1), we consider an aggregate literal $A$ occurring in $\mathcal{P}$. By previous considerations, we have that $A$ is true with respect to $I$ if and only if there is a conjunct in $trm(A)$ which is true with respect to $I$. Thus, (1) holds. □

We are now ready to relate our well-founded operator with the one provided by Pelov et al. (2007).

**Theorem 24** *For the class of $\mathrm{LP}_{m,a}^{\mathcal{A}}$ programs, the well-founded operator herein defined coincides with the one of Pelov et al. (2007; for both the ultimate and bound approximating aggregate semantics).*[3]

**Proof.** By Lemma 23, we already know that $\mathcal{W}_{\mathcal{P}}(I) = \mathcal{W}_{trm(\mathcal{P})}(I)$. We also have that $\mathcal{W}_{trm(\mathcal{P})}(I)$ coincides with the one in the work of Van Gelder et al. (1991) by Theorem 1 (since $trm(\mathcal{P})$ is a standard logic program). On the other hand, for both the ultimate and bound approximating aggregate semantics, the well-founded operators (as defined in Pelov et al., 2007) of $\mathcal{P}$ and $trm(\mathcal{P})$ coincide: This is a consequence of Theorem 1 in the work of Pelov et al. (2003), because the three-valued immediate consequence operators in the work of Pelov et al. (2003) and Pelov et al. (2007) coincide (see Definition 7 in Pelov et al., 2003 and Definition 7.5 in Pelov et al., 2007). Moreover, the well-founded operator of Pelov et al. (2007) coincides with the one in the work of Van Gelder et al. for standard logic programs, thereby obtaining the equality of the operators. □

The correspondence of the two well-founded semantics immediately follows from the theorem above. Indeed, the two well-founded models are defined as the fixpoints of the respective well-founded operators.

**Corollary 25** *The well-founded model herein defined and the one of Pelov et al. (2007; for both the ultimate and bound approximating aggregate semantics) coincide for $\mathrm{LP}_{m,a}^{\mathcal{A}}$ programs.*

As mentioned also earlier, by virtue of the above theorem and corollary, some of the results presented in this paper also follow from earlier results in the literature. In particular, Theorem 9, Theorem 16 and some of our complexity results follow from definitions and results of Pelov (2004) and Pelov et al. (2007).

---

3. Note that this operator is referred to as *stable revision operator* by Pelov et al. (2007).

## 8. Conclusion

In this paper we introduced a new notion of unfounded set for $LP_{m,a}^{\mathcal{A}}$ programs and analyzed a well-founded semantics for this language based on this notion. This semantics generalizes the traditional well-founded semantics for aggregate-free programs and also coincides with well-founded semantics for aggregate programs as defined by Pelov et al. (2007; the latter not being defined by means of a notion of unfounded set). We could also show that this semantics and its main operator $\mathcal{W}_{\mathcal{P}}$ have close ties with answer sets as defined by Faber et al. (2004, 2011), and can hence serve as approximations.

We proved that computing this semantics is a tractable problem. Indeed, the semantics is given by the least fixpoint of the well-founded operator $\mathcal{W}_{\mathcal{P}}$. The fixpoint is reached after a polynomial number of applications of the operator $\mathcal{W}_{\mathcal{P}}$ (with respect to the size of the input program), each of them requiring polynomial time. For showing that an application of $\mathcal{W}_{\mathcal{P}}$ is polynomial-time feasible, we have proved that evaluating monotone and antimonotone aggregate literals remains polynomial-time computable also for partial interpretations, since in this case only one of the possibly exponential extensions must be checked. For a monotone aggregate literal, this extension is obtained by falsifying each undefined literal, while for an antimonotone aggregate literal, each undefined literal is taken as true in the extension.

Motivated by these positive theoretical results, we have implemented the first system supporting a well-founded semantics for unrestricted $LP_{m,a}^{\mathcal{A}}$. Allowing for using monotone and antimonotone aggregate literals, the implemented prototype is ready for experimenting with the $LP_{m,a}^{\mathcal{A}}$ framework. The experiments conducted on the Attacks benchmark highlight the computational gains of a native implementation of aggregate constructs with respect to equivalent encodings in standard LP.

## Acknowledgments

Partly supported by Regione Calabria and EU under POR Calabria FESR 2007-2013 within the PIA project of DLVSYSTEM s.r.l., and by MIUR under the PRIN project LoDeN and under the PON project FRAME proposed by Atos Italia S.p.a.; we also thank the anonymous reviewers for their valuable comments.

## References

Alviano, M., Faber, W., & Leone, N. (2008). Compiling minimum and maximum aggregates into standard ASP. In Formisano, A. (Ed.), *Proceedings of the 23rd Italian Conference on Computational Logic (CILC 2008)*.

Baral, C. (2003). *Knowledge Representation, Reasoning and Declarative Problem Solving*. Cambridge University Press.

Brewka, G. (1996). Well-Founded Semantics for Extended Logic Programs with Dynamic Preferences. *Journal of Artificial Intelligence Research*, *4*, 19–36.

Calimeri, F., Faber, W., Leone, N., & Perri, S. (2005). Declarative and Computational Properties of Logic Programs with Aggregates. In *Nineteenth International Joint Conference on Artificial Intelligence (IJCAI-05)*, pp. 406–411.

Dell'Armi, T., Faber, W., Ielpa, G., Leone, N., & Pfeifer, G. (2003). Aggregate Functions in DLV. In de Vos, M., & Provetti, A. (Eds.), *Proceedings ASP03 - Answer Set Programming: Advances in Theory and Implementation*, pp. 274–288, Messina, Italy. Online at `http://CEUR-WS.org/Vol-78/`.

Denecker, M., Pelov, N., & Bruynooghe, M. (2001). Ultimate Well-Founded and Stable Model Semantics for Logic Programs with Aggregates. In Codognet, P. (Ed.), *Proceedings of the 17th International Conference on Logic Programming*, pp. 212–226. Springer Verlag.

Dix, J., & Osorio, M. (1997). On Well-Behaved Semantics Suitable for Aggregation. In *Proceedings of the International Logic Programming Symposium (ILPS '97)*, Port Jefferson, N.Y.

Eiter, T., Gottlob, G., & Mannila, H. (1997). Disjunctive Datalog. *ACM Transactions on Database Systems, 22*(3), 364–418.

Faber, W. (2005). Unfounded Sets for Disjunctive Logic Programs with Arbitrary Aggregates. In Baral, C., Greco, G., Leone, N., & Terracina, G. (Eds.), *Logic Programming and Nonmonotonic Reasoning — 8th International Conference, LPNMR'05, Diamante, Italy, September 2005, Proceedings*, Vol. 3662 of *Lecture Notes in Computer Science*, pp. 40–52. Springer Verlag.

Faber, W., Leone, N., & Pfeifer, G. (2004). Recursive aggregates in disjunctive logic programs: Semantics and complexity. In Alferes, J. J., & Leite, J. (Eds.), *Proceedings of the 9th European Conference on Artificial Intelligence (JELIA 2004)*, Vol. 3229 of *Lecture Notes in AI (LNAI)*, pp. 200–212. Springer Verlag.

Faber, W., Leone, N., & Pfeifer, G. (2011). Semantics and complexity of recursive aggregates in answer set programming. *Artificial Intelligence, 175*(1), 278–298. Special Issue: John McCarthy's Legacy.

Ferraris, P. (2005). Answer Sets for Propositional Theories. In Baral, C., Greco, G., Leone, N., & Terracina, G. (Eds.), *Logic Programming and Nonmonotonic Reasoning — 8th International Conference, LPNMR'05, Diamante, Italy, September 2005, Proceedings*, Vol. 3662 of *Lecture Notes in Computer Science*, pp. 119–131. Springer Verlag.

Ferraris, P. (2011). Logic programs with propositional connectives and aggregates. *ACM Transactions on Computational Logic, 12*(4). In press.

Gebser, M., Kaufmann, B., Neumann, A., & Schaub, T. (2007). Conflict-driven answer set solving. In *Twentieth International Joint Conference on Artificial Intelligence (IJCAI-07)*, pp. 386–392. Morgan Kaufmann Publishers.

Gelfond, M. (2002). Representing Knowledge in A-Prolog. In Kakas, A. C., & Sadri, F. (Eds.), *Computational Logic. Logic Programming and Beyond*, Vol. 2408 of *LNCS*, pp. 413–451. Springer.

Gelfond, M., & Lifschitz, V. (1991). Classical Negation in Logic Programs and Disjunctive Databases. *New Generation Computing, 9*, 365–385.

Gottlob, G., Leone, N., & Veith, H. (1999). Succinctness as a Source of Expression Complexity. *Annals of Pure and Applied Logic, 97*(1–3), 231–260.

Kemp, D. B., & Stuckey, P. J. (1991). Semantics of Logic Programs with Aggregates. In Saraswat, V. A., & Ueda, K. (Eds.), *Proceedings of the International Symposium on Logic Programming (ISLP'91)*, pp. 387–401. MIT Press.

Leone, N., Pfeifer, G., Faber, W., Eiter, T., Gottlob, G., Perri, S., & Scarcello, F. (2006). The DLV System for Knowledge Representation and Reasoning. *ACM Transactions on Computational Logic*, *7*(3), 499–562.

Liu, L., Pontelli, E., Son, T. C., & Truszczynski, M. (2010). Logic programs with abstract constraint atoms: The role of computations. *Artificial Intelligence*, *174*(3–4), 295–315.

Liu, L., & Truszczyński, M. (2006). Properties and applications of programs with monotone and convex constraints. *Journal of Artificial Intelligence Research*, *27*, 299–334.

Manna, M., Ruffolo, M., Oro, E., Alviano, M., & Leone, N. (2011). The HiLeX System for Semantic Information Extraction. *Transactions on Large-Scale Data and Knowledge-Centered Systems*. Springer Berlin/Heidelberg, To appear.

Manna, M., Ricca, F., & Terracina, G. (2011). Consistent Query Answering via ASP from Different Perspectives: Theory and Practice. *Theory and Practice of Logic Programming*, To appear.

Marek, V. W., & Truszczyński, M. (2004). Logic programs with abstract constraint atoms. In *Proceedings of the Nineteenth National Conference on Artificial Intelligence (AAAI 2004)*, pp. 86–91. AAAI Press / The MIT Press.

McCarthy, J. (1959). Programs with Common Sense. In *Proceedings of the Teddington Conference on the Mechanization of Thought Processes*, pp. 75–91. Her Majesty's Stationery Office.

McCarthy, J. (1980). Circumscription — a Form of Non-Monotonic Reasoning. *Artificial Intelligence*, *13*(1–2), 27–39.

McCarthy, J. (1986). Applications of Circumscription to Formalizing Common-Sense Knowledge. *Artificial Intelligence*, *28*(1), 89–116.

McCarthy, J. (1990). *Formalization of Common Sense, papers by John McCarthy edited by V. Lifschitz*. Ablex.

McCarthy, J., & Hayes, P. J. (1969). Some Philosophical Problems from the Standpoint of Artificial Intelligence. In Meltzer, B., & Michie, D. (Eds.), *Machine Intelligence 4*, pp. 463–502. Edinburgh University Press. reprinted in (McCarthy, 1990).

McDermott, D. V. (1982). Non-Monotonic Logic II: Nonmonotonic Modal Theories. *Journal of the ACM*, *29*(1), 33–57.

McDermott, D. V., & Doyle, J. (1980). Non-Monotonic Logic I. *Artificial Intelligence*, *13*(1–2), 41–72.

Minsky, M. (1975). A Framework for Representing Knowledge. In Winston, P. H. (Ed.), *The Psychology of Computer Vision*, pp. 211–277. McGraw-Hill.

Moore, R. C. (1985). Semantical Considerations on Nonmonotonic Logic. *Artificial Intelligence*, *25*(1), 75–94.

Osorio, M., & Jayaraman, B. (1999). Aggregation and Negation-As-Failure. *New Generation Computing*, *17*(3), 255–284.

Pelov, N. (2004). *Semantics of Logic Programs with Aggregates*. Ph.D. thesis, Katholieke Universiteit Leuven, Leuven, Belgium.

Pelov, N., Denecker, M., & Bruynooghe, M. (2003). Translation of Aggregate Programs to Normal Logic Programs. In de Vos, M., & Provetti, A. (Eds.), *Proceedings ASP03 - Answer Set Programming: Advances in Theory and Implementation*, pp. 29–42, Messina, Italy. Online at `http://CEUR-WS.org/Vol-78/`.

Pelov, N., Denecker, M., & Bruynooghe, M. (2004). Partial stable models for logic programs with aggregates. In *Proceedings of the 7th International Conference on Logic Programming and Non-Monotonic Reasoning (LPNMR-7)*, Vol. 2923 of *Lecture Notes in AI (LNAI)*, pp. 207–219. Springer.

Pelov, N., Denecker, M., & Bruynooghe, M. (2007). Well-founded and Stable Semantics of Logic Programs with Aggregates. *Theory and Practice of Logic Programming*, *7*(3), 301–353.

Pelov, N., & Truszczyński, M. (2004). Semantics of disjunctive programs with monotone aggregates - an operator-based approach. In *Proceedings of the 10th International Workshop on Non-monotonic Reasoning (NMR 2004), Whistler, BC, Canada*, pp. 327–334.

Reiter, R. (1980). A Logic for Default Reasoning. *Artificial Intelligence*, *13*(1–2), 81–132.

Ricca, F., Alviano, M., Dimasi, A., Grasso, G., Ielpa, S. M., Iiritano, S., Manna, M., & Leone, N. (2010). A Logic-Based System for e-Tourism. *Fundamenta Informaticae*. IOS Press, *105*(1–2), 35–55.

Ricca, F., Grasso, G., Alviano, M., Manna, M., Lio, V., Iiritano, S., & Leone, N. (2011). Team-building with Answer Set Programming in the Gioia-Tauro Seaport. *Theory and Practice of Logic Programming*. Cambridge University Press, To appear.

Ross, K. A., & Sagiv, Y. (1997). Monotonic Aggregation in Deductive Databases. *Journal of Computer and System Sciences*, *54*(1), 79–97.

Simons, P., Niemelä, I., & Soininen, T. (2002). Extending and Implementing the Stable Model Semantics. *Artificial Intelligence*, *138*, 181–234.

Son, T. C., & Pontelli, E. (2007). A Constructive semantic characterization of aggregates in answer set programming. *Theory and Practice of Logic Programming*, *7*, 355–375.

Son, T. C., Pontelli, E., & Tu, P. H. (2007). Answer Sets for Logic Programs with Arbitrary Abstract Constraint Atoms. *Journal of Artificial Intelligence Research*, *29*, 353–389.

Swift, T., & Warren, D. S. (2010). XSB: Extending prolog with tabled logic programming. *Computing Research Repository (CoRR)*, *abs/1012.5123*.

Tarski, A. (1955). A lattice-theoretical fixpoint theorem and its applications. *Pacific J. Math*, *5*, 285–309.

Truszczyński, M. (2010). Reducts of propositional theories, satisfiability relations, and generalizations of semantics of logic programs. *Artificial Intelligence*, *174*, 1285–1306.

Ullman, J. D. (1989). *Principles of Database and Knowledge Base Systems*. Computer Science Press.

Van Gelder, A. (1992). The Well-Founded Semantics of Aggregation. In *Proceedings of the Eleventh Symposium on Principles of Database Systems (PODS'92)*, pp. 127–138. ACM Press.

Van Gelder, A., Ross, K. A., & Schlipf, J. S. (1991). The Well-Founded Semantics for General Logic Programs. *Journal of the ACM*, *38*(3), 620–650.

Wittocx, J., Mariën, M., & Denecker, M. (2008). The IDP system: A model expansion system for an extension of classical logic. In Denecker, M. (Ed.), *Proceedings of the 2nd Workshop on Logic and Search, Computation of Structures from Declarative Descriptions (LaSh'08)*, pp. 153–165.